\documentclass[prd,twocolumn,notitlepage,superscriptaddress,nofootinbib]{revtex4-2}

\usepackage{amsmath}
\usepackage{amsfonts}
\usepackage{graphicx}
\usepackage[utf8]{inputenc}
\usepackage{slashed}

\usepackage{hyperref}
\hypersetup{colorlinks, linkcolor = [rgb]{0,0.0,0.75}, citecolor = [rgb]{0,0.0,0.75}, urlcolor = [rgb]{0,0.0,0.75}}

\makeatletter
\g@addto@macro\bfseries{\boldmath}
\makeatother

\newcommand{\MSb}{\overline{\text{MS}}}
\newcommand{\LambdaQCD}{\Lambda_{\textrm{QCD}}}
\newcommand{\numax}{\nu_{\textrm{max}}}
\newcommand{\zmax}{z_{\textrm{max}}}
\newcommand{\BGp}{{\textrm{BG}}_{\textrm{precond}}}

\begin{document}

\title{Flavor non-singlet parton distribution functions \\ from lattice QCD at physical quark masses \\ via the pseudo-distribution approach}

\author{Manjunath Bhat}
\affiliation{Faculty of Physics, Adam Mickiewicz University, ul.\ Uniwersytetu Pozna\'nskiego 2, 61-614 Poznań, Poland}
\author{Krzysztof~Cichy}
\affiliation{Faculty of Physics, Adam Mickiewicz University, ul.\ Uniwersytetu Pozna\'nskiego 2, 61-614 Poznań, Poland}
\author{Martha~Constantinou}
\affiliation{Temple University, 1925 N.\ 12th Street, Philadelphia, PA 19122-1801, USA}
\author{Aurora~Scapellato}
\affiliation{Faculty of Physics, Adam Mickiewicz University, ul.\ Uniwersytetu Pozna\'nskiego 2, 61-614 Poznań, Poland}

\date{\today}

\begin{abstract}
One of the great challenges of QCD is to determine the partonic structure of the nucleon from first principles. In this work, we provide such a determination of the flavor non-singlet ($u-d$) unpolarized parton distribution function (PDF), utilizing the non-perturbative formulation of QCD on the lattice.
We apply Radyushkin's pseudo-distribution approach to lattice results obtained using simulations with the light quark mass fixed to its physical value; this is the first ever attempt for this approach directly at the physical point.
The extracted coordinate-space matrix elements are used to find the relevant physical Ioffe time distributions from a matching procedure.
The full Bjorken-$x$ dependence of PDFs is resolved using several reconstruction methods to tackle the ill-conditioned inverse problem encountered when using discrete lattice data. 
We consider both the valence distribution $q_v$ and the combination with antiquarks $q_v+2\bar{q}$, related to, respectively, the real and imaginary part of extracted matrix elements.
Good agreement is found with PDFs from global fits already within statistical uncertainties and it is further improved by quantifying several systematic effects.
The results presented here are the first ever \emph{ab initio} determinations of PDFs fully consistent with global fits in the whole $x$-range.
Thus, they pave the way to investigating a wider class of partonic distributions, such as e.g.\ singlet PDFs and generalized parton distributions. Therefore, essential and yet missing first-principle insights can be achieved, complementing the rich experimental programs dedicated to the structure of the nucleon.
\end{abstract}

\maketitle

\section{Introduction}
\noindent Despite the fact that the nucleon is the main building block of visible matter and is responsible for almost all of the mass of the visible Universe, it is only now that several aspects of its internal structure are beginning to be thoroughly explored. The wealth of data from present-day experiments, e.g.\ from the Large Hadron Collider and the Jefferson Laboratory 6 and 12 GeV programs, allows us to unravel many details that so far eluded any insight.
Moreover, the planned and recently approved Electron-Ion Collider at the Brookhaven National Laboratory, that will start operation in around ten years, is oriented at answering important questions about the nucleon structure, such as of the origin of the proton mass, the spin distribution and the role of gluons \cite{NAP25171}.
Along with the experimental efforts, there is constant progress in the theoretical understanding, based on exhaustive analyses of empirical data and various approaches to describe the physics of the strong dynamics of partons (quarks and gluons), governed by the theory of quantum chromodynamics (QCD).
One of the most important tools is perturbation theory, which is, however, limited to high energy scales, at which the perturbative expansion can evince convergence.
Meanwhile, it is clear that all energy scales contribute to the dynamics of the nucleon and hence the description of the non-perturbative aspects is of essential importance.
This can take the form of phenomenological models, which have led to important successes.
However, a truly \emph{ab initio} knowledge can still be, in principle, extracted directly from the QCD Lagrangian. 
The most successful non-perturbative approach to QCD is to formulate it on a discrete spacetime grid, the lattice.
This leads to a regularization of the QCD path integral and results in multidimensional integrals that can be evaluated numerically, usually with Monte Carlo simulations.
Such computations, however, are necessarily done in Euclidean spacetime, which poses a fundamental problem for partonic physics, most naturally formulated in terms of light-cone correlations.
The latter cannot be accessed in Euclidean lattice QCD (LQCD) and thus, the information on the nucleon structure from LQCD has been limited for many years.
Most of the insights were obtained from lattice calculations of moments of parton distribution functions (PDFs) and generalized parton distributions (GPDs).
In principle, the full distributions can be reconstructed from a sufficiently large number of their moments, as obtained from LQCD. However, in practice, the computations are limited to only the lowest 2-3 moments.
For higher moments, the breaking of rotational symmetry implied by the lattice leads to unavoidable power-divergent mixings with lower-dimensional operators and what is more, the signal-to-noise ratio for such higher moments is quickly decaying.
For recent calculations of the low moments by the Extended Twisted Mass Collaboration (ETMC), we refer the interested reader to Refs.~\cite{Abdel-Rehim:2015owa,Oehm:2018jvm,Alexandrou:2018sjm,Alexandrou:2019ali,Alexandrou:2019brg,Alexandrou:2019olr}.

Extending the calculations of moments to the full $x$-dependence of the partonic distributions has been a subject of intense studies over the years.
Even though the early proposals date back to the previous century, the recent revival of this topic, one that brought considerable progress to this field, came with the proposal of Ji \cite{Ji:2013dva} to calculate so-called quasi-distributions.
They are related to the desired light-cone distributions, but computable on a Euclidean lattice.
The underlying idea is to replace the light-cone correlations in the definition of a distribution by spatial ones between boosted nucleon states. One can then exploit the fact that the obtained quasi-distributions share the infrared physics with their light-cone counterparts.
As a consequence, their difference is in the ultraviolet and can be matched using perturbation theory, utilizing the so-called Large Momentum Effective Theory (LaMET) \cite{Ji:2014gla}.
Ji's proposal sparked a huge tide of theoretical and numerical efforts to understand crucial aspects of this approach, such as renormalizability and  appropriate renormalization prescriptions, matching, nucleon mass corrections, higher-twist effects, finite volume effects, as well as to extract the distributions in various setups, see e.g.\ Refs.~\cite{Xiong:2013bka,Lin:2014zya,Gamberg:2014zwa,Alexandrou:2015rja,Gamberg:2015opc,Jia:2015pxx,Chen:2016utp,Chen:2016fxx,Alexandrou:2016jqi,Bacchetta:2016zjm,Briceno:2017cpo,Ishikawa:2017faj,Ji:2017oey,Constantinou:2017sej,Alexandrou:2017huk,Ji:2017rah,Wang:2017qyg,Green:2017xeu,Stewart:2017tvs,Broniowski:2017wbr,Broniowski:2017gfp,Izubuchi:2018srq,Alexandrou:2018pbm,Chen:2018fwa,Briceno:2018lfj,Spanoudes:2018zya,Radyushkin:2018nbf,Karpie:2018zaz,Liu:2018uuj,Lin:2018qky,Jia:2018qee,Bhattacharya:2018zxi,Alexandrou:2018eet,Braun:2018brg,Karpie:2019eiq,Alexandrou:2019lfo,Bhattacharya:2019cme,Chen:2019lcm,Izubuchi:2019lyk,Cichy:2019ebf,Son:2019ghf,Chai:2020nxw,Green:2020xco,Ji:2020baz,Bhattacharya:2020cen,Braun:2020ymy,Bhattacharya:2020xlt,Fan:2020nzz,Chen:2020arf,Zhang:2020dbb,Chen:2020iqi,Bhattacharya:2020jfj,Chen:2020ody,DelDebbio:2020cbz,Gao:2020ito,Alexandrou:2020tqq,Ji:2020brr,Alexandrou:2020zbe,Bringewatt:2020ixn,Liu:2020rqi,Alexandrou:2020qtt}.

Quasi-distributions can be thought of as generalization of the light-cone ones to the finite-momentum frame.
As shown by Radyushkin in a series of papers \cite{Radyushkin:2016hsy,Radyushkin:2017cyf,Radyushkin:2017lvu,Radyushkin:2017sfi,Radyushkin:2018cvn,Radyushkin:2018nbf,Radyushkin:2019owq,Radyushkin:2019mye},
the same matrix elements that specify a quasi-distribution can also be used to define another generalization of its light-cone counterpart, the so-called pseudo-distribution.
These matrix elements can be viewed as functions of two Lorentz invariants.
The first one is the spacetime interval $z^2$, where $z^\mu$ can be taken as $(0,0,0,z_3)$ and describes the separation between the quarks in the inserted operator (containing the Wilson line to guarantee gauge invariance).
The other Lorentz invariant is the product $\nu\equiv-p\cdot z$ (with $p^\mu$ being the nucleon boost) and the variable $\nu$ is called the Ioffe time \cite{Ioffe:1969kf}.
The matrix elements written as the function of $z^2$ and $\nu$ are called Ioffe-time distributions (ITDs).

The underlying difference between quasi- and pseudo-distributions is that the former are defined as the Fourier transform of the matrix elements in $z_3$, while to obtain the latter, one takes the transform in the Ioffe time.
This has far-reaching consequences and makes the approaches inequivalent, even though they can be computed from the same matrix elements.
In particular, pseudo-distributions have the canonical support in the Bjorken-$x$ fraction, $-1\leq x\leq1$, as opposed to quasi ones that can be non-zero also for $x$ outside of this range.
The matching of pseudo-distributions to the light-cone frame \cite{Radyushkin:2018cvn,Zhang:2018ggy,Izubuchi:2018srq,Radyushkin:2018nbf} is peformed at the level of ITDs. As we shall see below, the matching is numerically a smaller effect than for quasi-PDFs. In particular, it depends less significantly on the region of $x$.
It is also an important difference that pseudo-distributions can fully utilize lattice data at all nucleon boosts, i.e.\ they contain physical information also from low momenta, effectively at small Ioffe times.
Thus, the matching of ITDs is not based on LaMET, but still on a factorization of the finite-momentum distribution into a light-cone one and a perturbatively computable matching coefficient.
The pseudo-distribution approach to extract the valence unpolarized PDF of the nucleon was explored first in the quenched setup \cite{Orginos:2017kos} and recently it was extended to include the effects of dynamical quarks at non-physical masses \cite{Joo:2019jct} and to investigate the PDFs of the pion \cite{Joo:2019bzr}.
Even more recently, it was also applied in a setup with light quark mass close to, but not yet directly at its physical value, with the corresponding pion mass of 170 MeV \cite{Joo:2020spy}.
In addition, moments of ITDs were computed \cite{Karpie:2018zaz} and the general issue of reconstructing distributions from ITDs under incomplete Fourier transforms was analyzed in related studies by the same group \cite{Karpie:2019eiq} and most recently, the pseudo-PDF data were used to reconstruct PDFs using neural networks \cite{DelDebbio:2020rgv}.
.

Apart from approaches based on ITDs, we also mention other proposed methods that can lead to determinations of the full $x$-dependence of partonic distributions.
Instead of using a Wilson line to ensure gauge invariance, they can employ auxiliary propagators, of fictitious scalar \cite{Aglietti:1998ur}, heavy \cite{Detmold:2005gg} or light quarks \cite{Braun:2007wv,Ma:2014jla,Ma:2017pxb}.
Approaches based on the hadronic tensor also exist \cite{Liu:1993cv,Chambers:2017dov}.
Similarly to quasi- and pseudo-distributions, these methods are also intensely investigated by various groups, see e.g.\ Refs.~\cite{Bali:2018spj,Detmold:2018kwu,Sufian:2019bol,Liang:2019frk,Sufian:2020vzb,Can:2020sxc}. 
An extensive review of these efforts, with emphasis on the most investigated approach of quasi-distributions, can be found in Ref.\ \cite{Cichy:2018mum,Constantinou:2020pek}.
Recently, a review devoted to the principles and various applications of LaMET also appeared \cite{Ji:2020ect}.

In this paper, we apply the pseudo-distribution approach for the first time to lattice data obtained with light quark mass fixed to its physical value.
We always consider the flavor non-singlet combination $u-d$.
We provide an extraction of two kinds of non-singlet distributions -- the valence one, denoted here by $q_v(x)$, and its combination with antiquarks $q_v(x)+2\bar{q}(x)$.
Taking their linear combinations, we also show results for the full distribution, $q(x)=q_v(x)+\bar{q}(x)$ and for the sea quark PDF, $q_s(x)=\bar{q}(x)$.
We explore several systematic effects inherited in lattice computations and we address the issue of reconstructing the PDFs from ITDs, subject to an ill-defined inverse problem.
We show that this leads ultimately to full consistency between all considered distributions and the corresponding ones from global fits, in the whole range of Bjorken-$x$.
Even though at the present stage full quantification of all systematics is not yet possible, 
the results obtained in this work are unambiguously optimistic and demonstrate the potential of the techniques for an issue that was for many years thought to be too difficult for lattice QCD calculations.

The outline of the remainder of the paper is the following.
In Sec.~\ref{sec:theory}, we discuss theoretical principles of pseudo-PDFs and practical aspects of their computations.
Then, we present the lattice details of the calculation in Sec.~\ref{sec:lattice}.
In Sec.~\ref{sec:results}, we show our results for the pseudo-distributions and their matching to light-cone ITDs and we compare different methods for the reconstruction of PDFs.
We also discuss systematic effects from the choice of the Ioffe time range, the value of the strong coupling constant and from other sources and in Sec.~\ref{sec:final}, we present our final PDFs. 
We also show results for the low moments obtained from polynomial fits to fixed-$z^2$ ITDs and from integrating the final reconstructed distributions.
Finally, we conclude and discuss future prospects in Sec.~\ref{sec:summary}.

\section{Theoretical setup and analysis techniques}
\label{sec:theory}

We start by summarizing the relevant steps in the procedure leading from lattice-extracted matrix elements (pseudo-ITDs) to the light-cone distributions, ITDs and PDFs after a suitable reconstruction.
We refer the Reader to the review of Ref.~\cite{Radyushkin:2019owq} for an extensive discussion on the theoretical principles and properties of pseudo-distributions.

The underlying matrix elements computed on the lattice are defined (in Euclidean spacetime) as
\begin{equation}
\label{eq:ME}
\mathcal{M}(\nu,z^2)=\langle P\vert \, \overline{\psi}(0,z)\,\gamma_0 W(z,0)\,\psi(0,0)\,\vert P\rangle\, ,
\end{equation}
where $\vert P\rangle$ is a boosted nucleon state with four-momentum $P_\mu {=} (P_0, 0, 0, P_3)$ and $W(z,0)$ is a straight Wilson line. The Wilson line is chosen along the direction of the boost and has length $z$ (i.e.\ we take $z^\mu=(0,0,0,z)$ and $z$ will henceforth refer to the length of $z^\mu$). With this choice of kinematics, the Ioffe time $\nu=P_3z$.
The Dirac structure $\gamma_0$ leads to a faster convergence to the light-cone ITD, as compared to another plausible choice of the $\gamma_3$ structure \cite{Radyushkin:2016hsy}. It was also found that $\gamma_0$ avoids a finite mixing with the twist-3 scalar operator due to the breaking of chiral symmetry by some fermionic discretizations \cite{Constantinou:2017sej}.

The matrix element defined by Eq.~(\ref{eq:ME}) exhibits two kinds of divergences: standard  logarithmic one and a power divergence related to the Wilson line.
However, it has been shown to be multiplicatively renormalizable to all orders in perturbation theory \cite{Ishikawa:2017faj,Ji:2017oey}.
In Ref.~\cite{Orginos:2017kos}, it was suggested that the divergences can be canceled by forming a double ratio with zero-momentum and local ($z=0$) matrix elements:
\begin{equation}
\label{eq:reduced}
\mathfrak{M}(\nu,z^2) = \frac{\mathcal{M}(\nu,z^2)\,/\,\mathcal{M}(\nu,0)}
{\mathcal{M}(0,z^2)\,/\,\mathcal{M}(0,0)}.
\end{equation}
We follow this renormalization procedure and we refer to $\mathfrak{M}(\nu,z^2)$ as reduced matrix elements or pseudo-ITDs.
It is expected, although not proven, that the double ratio not only removes the divergences, but also part of the higher-twist contamination, which is generically of $\mathcal{O}(z^2\LambdaQCD^2)$ \cite{Orginos:2017kos}.
The double ratio defines a renormalization scheme where the renormalization scale is proportional to the inverse length of the Wilson line.

To get from pseudo-ITDs to light-cone ITDs and finally to light-cone PDFs, a matching procedure is required, similiarly to the quasi-PDFs case. 
The reduced matrix elements, defined at different scales $1/z$, need to be evolved to a common scale, $1/z'$, and it is desirable to also convert them to the renormalization scheme commonly used for PDFs, the $\MSb$ scheme, where its renormalization scale will be denoted by $\mu$.
The full matching equation, to one-loop order in perturbation theory, reads \cite{Radyushkin:2018cvn,Zhang:2018ggy,Izubuchi:2018srq,Radyushkin:2018nbf}:
\begin{eqnarray}
\mathfrak{M}(\nu,z^2) &=& Q(\nu,\mu^2) + \frac{\alpha_s C_F}{2\pi} \int_0^1 du \\
&\times& \left[ \ln \left(z^2\mu^2 \frac{e^{2\gamma_E +1}}4\right) B(u)
+ L(u) \right] Q(u\nu,\mu^2),\nonumber
\end{eqnarray}
where $Q(\nu,\mu^2)$ is the $\MSb$-scheme light-cone ITD and the functions convoluted with $Q$ are
\begin{equation}
\label{eq:B}
B(u)=\left[\frac{1+u^2}{1-u}\right]_+,
\end{equation}
\begin{equation}
\label{eq:L}
L(u) = \left[ 4 \frac{\ln(1-u)}{1-u} - 2(1-u) \right]_+
\end{equation}
with the plus prescription defined as
\begin{equation}
\int_0^1 \left[f(u)\right]_+ Q(u\nu) = \int_0^1 f(u)\left(Q(u\nu)-Q(\nu)\right). 
\end{equation}
The matching equation consists of two parts.
The part containing the kernel $B(u)$ evolves the pseudo-ITDs to a common scale $\mu$ and the part with $L(u)$ converts to the $\MSb$ scheme.

We invert the matching equation and to look separately into the effect of evolution and scheme conversion, we introduce intermediate evolved ITDs, $\mathfrak{M}'(\nu,z^2,\mu^2)$.
Thus,
\begin{eqnarray}
\label{eq:evol}
\mathfrak{M}'(\nu,z^2,\mu^2) &=& \mathfrak{M}(\nu,z^2) - \frac{\alpha_s C_F}{2\pi} \int_0^1 du \\
&\times& \ln \left(z^2\mu^2 \frac{e^{2\gamma_E +1}}4\right) 
 B(u) \mathfrak{M}(u\nu,z^2) \nonumber.
\end{eqnarray}
The evolved ITD has three arguments, the Ioffe time $\nu$, the common scale $\mu$ and the initial scale $z$.
In principle, its value should be independent of the initial scale and we will test this conjecture up to our statistical precision.
The scheme conversion then follows:
\begin{equation}
\label{eq:match}
Q(\nu,z^2,\mu^2) = \mathfrak{M}'(\nu,z^2,\mu^2) - \frac{\alpha_s C_F}{2\pi} \int_0^1 du  L(u)  \mathfrak{M}(u\nu,z^2),
\end{equation}
where again we will test the independence on the initial scale.
For the reconstruction of the final PDF, discussed below, we will average the matched ITDs $Q(\nu,z^2,\mu^2)$ for cases where a given Ioffe time is achieved by different combinations of $(P_3,z)$ and denote such an average by $Q(\nu,\mu^2)$.

The matching equations discussed above involve a convolution of a kernel function with reduced ITDs, i.e.\ one needs to access the latter for all continuous values of the Ioffe time from 0 to $\nu$.
We adopt two alternative approaches for the required interpolation at fixed $z^2$.
We either do a linear interpolation between the ITDs at neighboring boosts or we perform fits to the $\nu$-dependence that utilizes all boosts. The real part is then fitted to even powers of the Ioffe time,
\begin{equation}
\label{eq:fitRe}
{\rm Re}\,\mathfrak{M}(\nu,z^2)=1+c_2\nu^2+c_4\nu^4+\ldots,
\end{equation}
while the imaginary part to its odd powers:
\begin{equation}
\label{eq:fitIm}
{\rm Im}\,\mathfrak{M}(\nu,z^2)=c_1\nu+c_3\nu^3+\ldots.
\end{equation}
We compare both approaches at the level of matrix elements and the final reconstructed partonic distributions and show that both of them lead to totally consistent outcomes.
However, the fitting of the full boost dependence at fixed $z^2$ allows one to also extract PDF moments, as suggested in Ref.~\cite{Karpie:2018zaz}.
Moments of pseudo-PDFs are defined according to \cite{Karpie:2018zaz} 
\begin{equation}
\label{eq:moments_pseudo}
\mathfrak{M}_n(z^2)=(-i)^n \frac{\partial^n\mathfrak{M}(\nu,z^2)}{\partial \nu^n}\Bigg|_{\nu=0}
\end{equation}
and are, thus, related to the fitting coefficients $c_n$.
To extract moments of PDFs in the $\MSb$ scheme at a scale $\mu^2$, denoted here by $Q_n(\mu^2)$, one utilizes a matching equation \cite{Karpie:2018zaz,Joo:2020spy} that factorizes the pseudo-PDF moments into $\MSb$ moments and a perturbative coefficient expressing the Mellin moments of the ITDs matching kernel,
\begin{equation}
\label{eq:moments_matching}
\mathfrak{M}_n(z^2) = C_n(\mu^2z^2) Q_n(\mu^2)
 \end{equation}
up to $\mathcal{O}(z^2\LambdaQCD^2)$ higher-twist effects (HTE).
At one-loop order,
\begin{equation}
\label{eq:moments_coeff}
C_n(\mu^2 z^2) = 
1 - \frac{\alpha_s C_F}{2\pi} \left[\gamma_n \ln\left(z^2\mu^2\frac{e^{2\gamma_E +1}}{4}\right) + l_n\right], 
\end{equation}
with $\gamma_n$ being the moments of the kernel $B(u)$ of Eq.~(\ref{eq:B}), 
\begin{equation}
\gamma_n = \frac{1}{(n+1)(n+2)} - \frac{1}{2} - 2 \sum_{k=2}^{n+1}\frac{1}{k},
\end{equation} 
and $l_n$ the moments of $L(u)$ of Eq.~(\ref{eq:L}),
\begin{equation}
l_n = 2\left[ \left(\sum_{k=1}^n \frac{1}{k}\right)^2 + \sum_{k=1}^n \frac{1}{k^2} 
+\frac{1}{2} - \frac{1}{(n+1)(n+2)} \right].
\end{equation}

We now move on to a discussion on how to extract partonic distributions from the evolved and scheme-converted ITDs.
The ITDs, $Q(\nu,\mu^2)$, are related to PDFs, $q(x,\mu^2)$, by a Fourier transform in Ioffe time:
\begin{equation}
\label{eq:PDF2ITD}
Q(\nu,\mu^2) =\int_{-1}^1 dx \, e^{i\nu x} q(x,\mu^2),
\end{equation}
where the antiquark distribution for positive $x$ is $\bar{q}(x)=-q(-x)$.
Decomposing into real and imaginary parts and using this property, one obtains
\begin{eqnarray}
\label{eq:ReQ}
{\rm Re}\,Q(\nu,\mu^2) & =&\int_0^1 dx \cos(\nu x) \big(q(x,\mu^2)-\bar{q}(x,\mu^2)\big)\nonumber \\& =& \int_0^1 dx \cos(\nu x) q_v(x,\mu^2),
\end{eqnarray}
\begin{eqnarray}
\label{eq:ImQ}
{\rm Im}\,Q(\nu,\mu^2)& =&\int_0^1 dx \sin(\nu x) \big(q(x,\mu^2) + \bar{q}(x,\mu^2)\big)\nonumber\\& =& \int_0^1 dx \sin(\nu x) q_{v2s}(x,\mu^2),
\end{eqnarray}
which relate the valence distribution, $q_v=q-\bar{q}$, to the real part of the ITDs and the other non-singlet distribution involving two flavors, $q_{v2s}\equiv q_v+2\bar{q}=q+\bar{q}$, to the imaginary part of the ITDs.

All of the equations (\ref{eq:PDF2ITD})-(\ref{eq:ImQ}) involve a known left-hand side (reduced ITDs computed on the lattice and subject to the matching procedure) and integrals of a partonic distribution to be determined.
As discussed in detail in Ref.~\cite{Karpie:2018zaz}, such determination poses an inverse problem, related to the fact that inverse equations are ill-defined.
Namely, they involve an integral over continuous Ioffe time up to infinity, while on the lattice, one is necessarily restricted to a finite number of determinations of $Q(\nu)$ that cover only a finite range of Ioffe time, from 0 up to some $\numax$.
To reconstruct the distributions, we will follow three ways.
The inverse problem stems from having incomplete information and hence, solving it is not possible without additional assumptions.
There is an infinite number of possible assumptions to provide the missing information and thus, it is important to use as mild ones as possible, in order not to bias the reconstruction procedure.
We will perform naive Fourier transforms and we will use two additional ways of handling the inverse problem.
First, we will apply the Backus-Gilbert (BG) method \cite{BackusGilbert}, originally proposed to be used in PDF reconstruction in Ref.~\cite{Karpie:2018zaz}.
Second, we will perform reconstruction by fitting the matrix elements using a fitting ansatz for the light-cone PDF, as suggested in Ref.~\cite{Joo:2019jct}.

The BG method minimizes the variance of the solution to the inverse problem, i.e.\ it maximizes its stability with respect to variation of the data within their errors.
This variance minimization condition is a model-independent assumption that provides a unique distribution given a set of input ITDs.
For each value of Bjorken-$x$, the minimization condition defines a $d$-dimensional vector $\textbf{a}_K(x)$ (where $d$ is the number of available evaluations of the input ITD), which is an approximate inverse of the kernel function $\textbf{K}(x)$ (the cosine or the sine function, respectively for Eqs.~(\ref{eq:ReQ})-(\ref{eq:ImQ})), i.e.
\begin{equation}
\Delta(x-x')=\sum_{\nu}a_K(x)_\nu K(x')_\nu,    
\end{equation}
where $\textbf{K}(x')$ is taken as a $d$-dimensional vector with elements $K(x')_\nu=\cos(\nu x')$ or $K(x')_\nu=\sin(\nu x')$.
In the ideal case of $d\rightarrow\infty$ evaluations of the ITD spanning the infinite range of Ioffe times, thus defined vector $\textbf{a}_K(x)$ would lead to $\Delta(x-x')$ being the Dirac delta function $\delta(x-x')$.
In turn, for finite $d$, $\textbf{a}_K(x)$ leads to an approximation to the Dirac delta function with minimized width.
We find the vectors $\textbf{a}_K(x)$ from width minimization conditions, spelled out explicitly e.g.\ in Ref.~\cite{Karpie:2018zaz}, which yield
\begin{equation}
\textbf{a}_K(x)=\frac{\textbf{M}_K^{-1}(x)\,\textbf{u}_K}{\textbf{u}_K^T\,\textbf{M}_K^{-1}(x)\,\textbf{u}_K},    
\end{equation}
where the elements of the $d\times d$-dimensional matrix $\textbf{M}_K(x)$ are given by
\begin{equation}
M_K(x)_{\nu\nu'}=\int_0^1 dx'\,(x-x')^2 K(x')_\nu \, K(x')_{\nu'}+\rho\,\delta_{\nu\nu'}
\end{equation}
and of the $d$-dimensional vector $\textbf{u}_K$ by
\begin{equation}
u_{K\nu}=\int_0^1 dx'\,K(x')_\nu.    
\end{equation}
The parameter $\rho$ in $\textbf{M}_K(x)$ regularizes this matrix (Tikhonov regularization \cite{Tikhonov:1963}, see also Refs.~\cite{Ulybyshev:2017szp,Ulybyshev:2017ped,Karpie:2018zaz}), making it invertible.
The value of $\rho$ should be relatively small in order not to bias the result and not to  decrease the resolution of the method.
In our study, we find that $\rho=10^{-3}$ is a good compromise, with smaller values introducing large oscillations in the final distributions due to the presence of very small eigenvalues of $\textbf{M}_K(x)$.

Having found the vectors $\textbf{a}_K(x)$ (for both kernel functions), the distributions $q_v$ or $q_{v2s}$ are reconstructed as
\begin{equation}
q_{v/v2s}(x,\mu^2)=\sum_{\nu} a_K(x)_\nu \, {\rm Re}/{\rm Im}\, Q(\nu,\mu^2). 
\end{equation}
We will also consider a version of the BG procedure with preconditioning ($\BGp$), which can be applied in a case where a realistic guess of the solution of the inverse problem is available.
In this variant, the kernel function and the desired distribution are rescaled by a function $p(x)$, $\tilde{\textbf{K}}(x)=\textbf{K}(x)p(x)$ and $\tilde{q}(x)=q(x)/p(x)$.
Then, $\tilde{q}(x)$ embodies the deviation of the reconstructed distribution from $p(x)$.
The procedure of calculating the vectors $\textbf{a}_K(x)$ is unchanged, apart from using the rescaled kernel and taking into account the preconditioning function in the final reconstruction equation:
\begin{equation}
\label{eq:precond}
q_{v/v2s}(x,\mu^2)=\sum_{\nu} a_K(x)_\nu \,p(x)\, {\rm Re}/{\rm Im}\, Q(\nu,\mu^2). 
\end{equation}

The other reconstruction technique that we will use is to assume a functional form of a fitting ansatz for the light-cone PDF.
This is analogous to procedures used in phenomenological fits of PDFs from experimental data.
We will adhere to the simplest reasonable functional form that captures the expected limiting behaviors at low and large-$x$ for both $q_v$ and $q_{v2s}$, in the range $x\in(0,1)$:
\begin{equation}
\label{eq:ansatz}
q(x) = N x^a (1-x)^b,
\end{equation}
where the exponents $a,\,b$ are fitting parameters.
$N$ is fixed to $1/B(a+1,b+1)$ for $q_v$, where $B(x,y)$ is the Euler beta function, related to the gamma function, $B(x,y)=\Gamma(x)\Gamma(y)/\Gamma(x+y)$. 
This ensures the normalization of the valence distribution to 1.
In the case of $q_v+2\bar{q}$, $N$ is left as an additional fitting parameter.

The fits are performed minimizing the $\chi^2$ function defined as
\begin{equation}
\label{eq:chi2}
\chi^2=\sum_{\nu=0}^{\numax}\frac{Q(\nu,\mu^2)-Q_f(\nu,\mu^2)}{\sigma_Q^2(\nu,\mu^2)},
\end{equation}
where $\sigma_Q^2(\nu,\mu^2)$ is the statistical error of the light-cone ITD $Q(\nu,\mu^2)$.
$Q_f(\nu,\mu^2)$ is given by the cosine/sine Fourier transform of the fitting ansatz (\ref{eq:ansatz}), respectively for fits of the real/imaginary part of ITDs.
The fitting function is continuous and thus, this Fourier transform is not subject to any inverse problem.
We will refer to the values of $Q_f$ as ``fitted'' ITDs and they are a continuous function of the Ioffe time.
Note the fits depend on the maximum Ioffe time, $\numax$, and we will investigate different choices of this parameter and the sensitivity of the final PDF results to this choice.

\section{Lattice setup}
\label{sec:lattice}
The underlying matrix elements are the same as the ones used for the computation of quasi-PDFs.
Thus, we use the matrix elements of Eq.~(\ref{eq:ME}), calculated in Refs.~\cite{Alexandrou:2018pbm,Alexandrou:2019lfo}, corresponding to the Dirac structure $\gamma_0$ of the unpolarized PDF case.
For quasi-PDFs, the Fourier transform defining the distributions is performed at a fixed nucleon boost $P_3$ and the data of Refs.~\cite{Alexandrou:2018pbm,Alexandrou:2019lfo} concern the cases of $P_3=6\pi/L,\,8\pi/L,10\pi/L$, corresponding to 0.83 GeV, 1.11 GeV and 1.38 GeV in physical units, respectively.
In the pseudo-PDF approach, the Fourier transform is taken in Ioffe time and hence can profit also from data at low nucleon momenta.
Moreover, the double ratio that defines the reduced ITDs requires the knowledge of the zero-boost matrix elements.
Thus, for this work, we computed also the cases of $P_3=0,\,2\pi/L,4\pi/L$, i.e.\ 0, 0.28 GeV and 0.55 GeV.

For details of the computational techniques, we refer to the broad description in Ref.~\cite{Alexandrou:2019lfo}.
Here, we summarize the main aspects.
We use one ensemble of gauge field configurations with two degenerate flavors of maximally twisted mass fermions \cite{Frezzotti:2000nk,Frezzotti:2003ni} with a clover improvement \cite{Sheikholeslami:1985ij}, generated by the Extended Twisted Mass Collaboration (ETMC) \cite{Abdel-Rehim:2015pwa}.
The gauge action is Iwasaki-improved \cite{Iwasaki:2011np}.
The bare quark mass was tuned to approximately reproduce the physical value of the pion mass ($m_\pi=130.4(4)$ MeV) and the nucleon mass ($m_N=932(4)$ MeV) \cite{Alexandrou:2017xwd}.
The lattice spacing is $a=0.0938(2)(3)$ fm \cite{Alexandrou:2017xwd} and the lattice volume is $48^3\times96$ sites, which corresponds to a physical lattice extent of $L\approx4.5$ fm.

Twisted mass fermions at maximal twist evince automatic $\mathcal{O}(a)$ improvement of physical observables. 
However, the matrix elements of Eq.~(\ref{eq:ME}) are not in this category, apart from the local ($z=0$) case.
Thus, in general, the PDFs calculated in this work have leading cut-off effects linear in the lattice spacing.
As shown in Ref.~\cite{Green:2020xco}, maximal twist can remove some of the $\mathcal{O}(a)$ contributions, but still an explicit specific improvement program is necessary to fully eliminate them.
Likewise, an improvement program is needed also for other lattice discretizations, including ones preserving chiral symmetry \cite{Green:2020xco}.

\begin{table}[h!]
\begin{center}
\begin{tabular}{cc|cc}
\hline
$P_3$ & $P_3$ [GeV] & $N_{\rm confs}$ & $N_{\rm meas}$\\
\hline
$0$ & 0 & 20 & 320\\
$2\pi/L$ & 0.28 & 19 & 1824\\
$4\pi/L$ & 0.55 & 18 & 1728\\
$6\pi/L$ & 0.83 & 50 & 4800\\
$8\pi/L$ & 1.11 & 425 & 38250\\
$10\pi/L$ & 1.38 & 811 & 72990\\
\hline
\end{tabular}
\caption{The number of gauge field configurations ($N_{\rm confs}$) and the number of measurements ($N_{\rm meas}$) for each value of the nucleon boost used in this work.}
\label{tab:stat}
\end{center}
\end{table}

In Tab.~\ref{tab:stat}, we summarize the statistics for the computation of matrix elements. 
While the statistics for $P_3<0.83$ GeV can be further increased, we keep it lower than the three highest momenta. This is desirable, as we aimed at a similar statistical precision for all data.
At these low momenta, the signal-to-noise ratio is very favorable and thus, similar precision could be achieved already with the modest number of measurements reported in Tab.~\ref{tab:stat}.
For all momenta, we use a source-sink separation, $t_s$, of 12 lattice spacings (1.13 fm), which was shown \cite{Alexandrou:2019lfo} to suppress excited states effects to below the statistical precision of the data.
Since the contamination from excited states increases at larger momenta, the additional matrix element computations for this work can be safely assumed to be free from such effects when using $t_s=12a$, at our level of statistical precision.

\section{Results}
\label{sec:results}
\subsection{Bare and reduced matrix elements}
\label{sec:ME}
\begin{figure}[h!]
\begin{center}
    \includegraphics[width=0.48\textwidth]{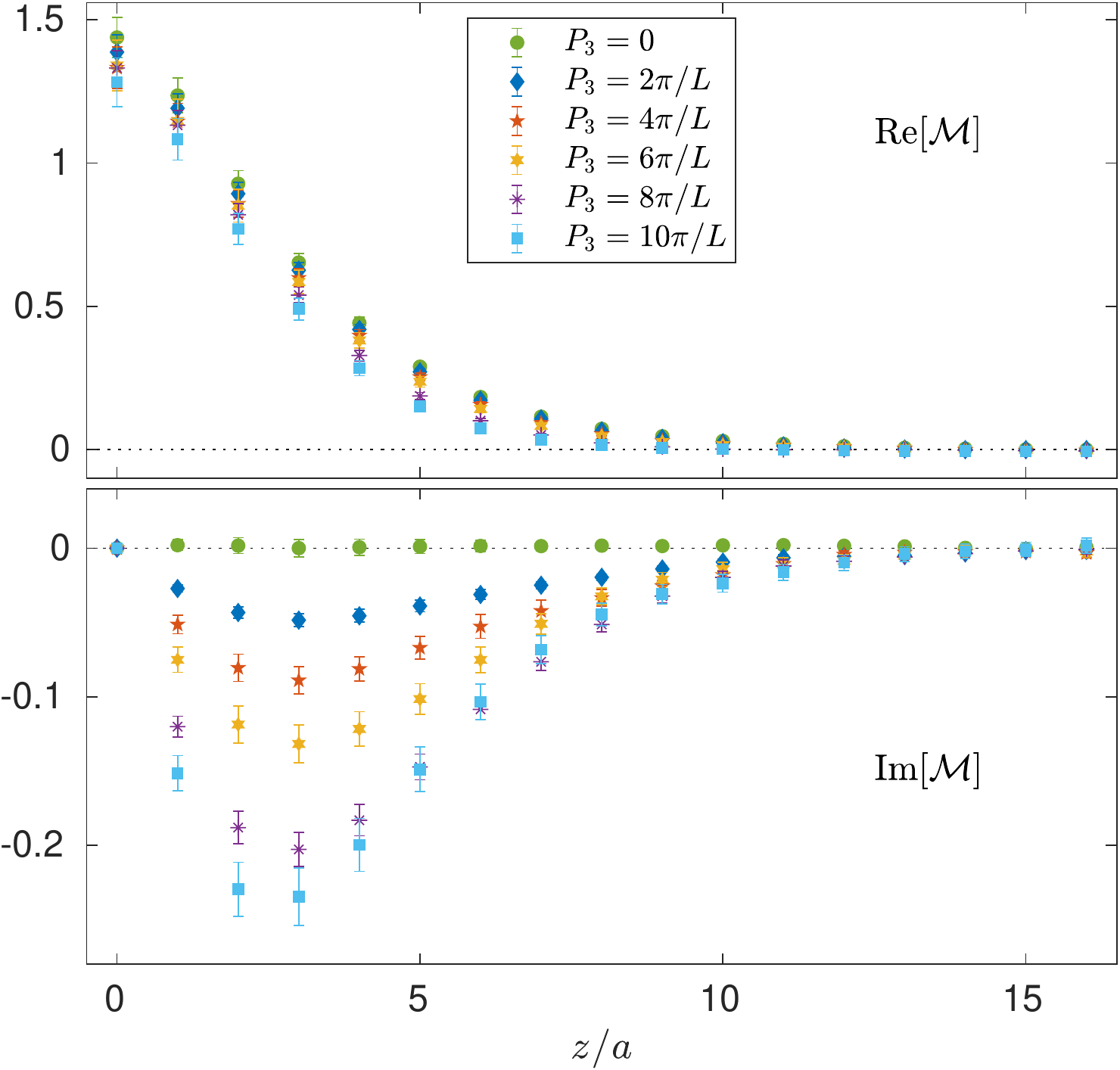}
\end{center}
\vspace*{-0.5cm} 
\caption{Real (top) and imaginary (bottom) part of the bare matrix elements ($\mathcal{M}(\nu,z^2)$ at fixed $P_3$) for the unpolarized PDF. Shown are all nucleon boosts: $P_3=0$ (green circles), $P_3=2\pi/L$ (blue rhombuses), $P_3=4\pi/L$ (red 5-stars), $P_3=6\pi/L$ (yellow 6-stars), $P_3=8\pi/L$ (purple asterisks), $P_3=10\pi/L$ (cyan squares).}
\label{fig:bare}
\end{figure}

We start by showing bare matrix elements as a function of $z/a$ for the 6 computed nucleon boosts, from 0 to 1.38 GeV, see Fig.~\ref{fig:bare}.
We observe that as the hadron momentum is increased, the distribution of the real part of the matrix elements becomes slightly narrower with respect to the length of the Wilson line, i.e.\ they decay to zero at smaller values of $z$.
At the same time, the imaginary part becomes more pronounced for larger momenta.
The local matrix element ($z=0$) is real and contains no divergence.
With the employed definition of the vector current, it is only subject to a normalization factor computed in Ref.~\cite{Alexandrou:2015sea}, $Z_V=0.7565(4)(19)$, and after multiplication by it, $\mathcal{M}(0,0)$ is compatible with 1 for all nucleon boosts.

\begin{figure}[h!]
\begin{center}
    \includegraphics[width=0.48\textwidth]{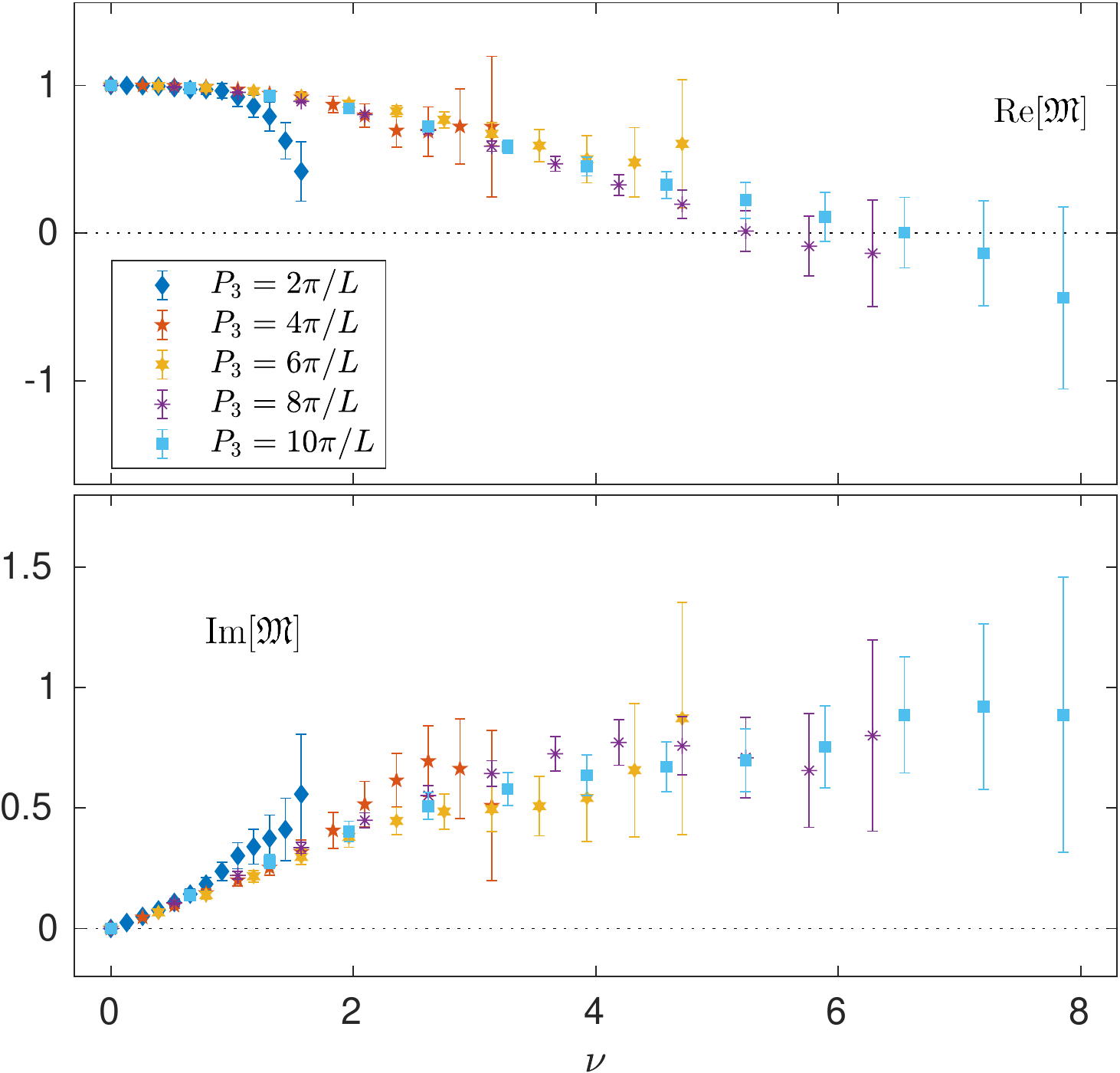}
\end{center}
\vspace*{-0.5cm} 
\caption{Real (top) and imaginary (bottom) part of the reduced matrix elements ($\mathfrak{M}(\nu,z^2)$ at fixed $P_3$) for the unpolarized PDF. Symbols are the same as used in Fig.~\ref{fig:bare}.}
\label{fig:reduced}
\end{figure}

We form the reduced matrix elements according to Eq.~(\ref{eq:reduced}) and plot them against the Ioffe time in Fig.~\ref{fig:reduced}, separately for each nucleon boost and for lengths of the Wilson line $z/a\in[0,12]$.
We observe that for values of $z/a$ smaller than approx.\ 8, reduced ITDs obtained from different combinations of $(P_3,z)$ that lead to the same Ioffe time $P_3z$ are compatible with each other within our statistical uncertainties. 
This suggests that as long as the difference in the scale $1/z$ is not too large, the scale-dependence of reduced ITDs is rather small.
The clearest deviations are observed in the real part for the lowest boost, $P_3=2\pi/L$, for $z/a>8$, where the scales $1/z$ correspond to around 250 MeV and below.

\begin{figure}[h!]
\begin{center}
    \includegraphics[width=0.48\textwidth]{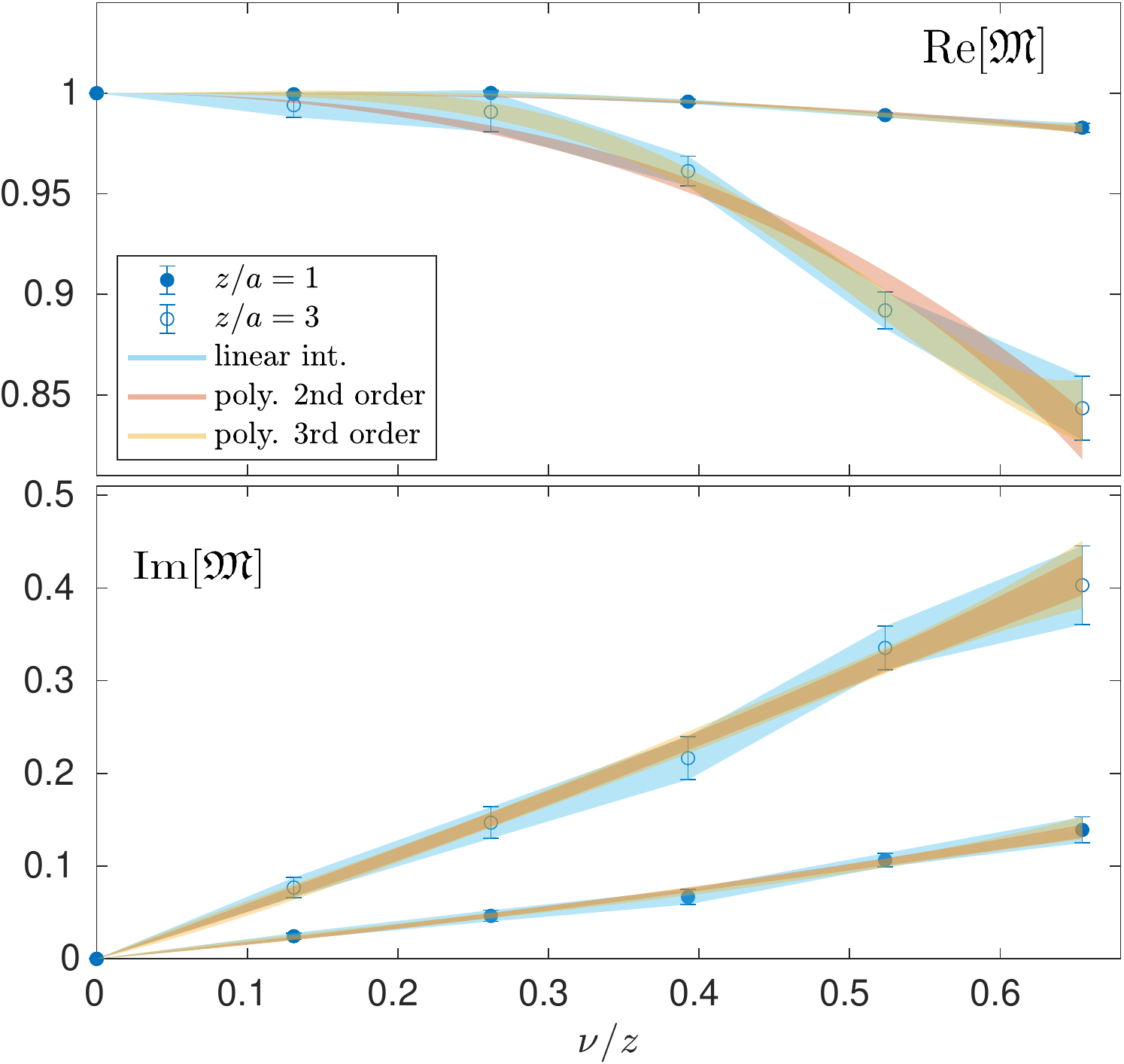}
\end{center}
\vspace*{-0.5cm} 
\caption{Illustration of our interpolation procedures to extract the real part (top) and the imaginary part (bottom) of reduced ITDs at fixed $z^2$ and arbitrary $\nu$. Shown are the actual lattice data at $(z/a)^2=1$ (closed symbols) and $(z/a)^2=9$ (open symbols), with linear interpolations between the neighboring points (blue bands) and second (red bands) and third order (orange bands) polynomial fits to the full Ioffe time dependence. The $\chi^2/{\rm dof}$ values of the polynomial fits are for the real part: 0.97/0.36 (2nd/3rd order, $z/a=1$), 1.67/0.61 ($z/a=3$) and for the imaginary part: 0.33/0.47 ($z/a=1$), 0.37/0.52 ($z/a=3$).}
\label{fig:interpolME}
\end{figure}

Before we move on to evolution and scheme conversion of the ITDs, we discuss our interpolation procedure at fixed $z^2$, which provides the reduced matrix elements at arbitrary, continuous values of the Ioffe time, needed to perform the matching.
In Fig.~\ref{fig:interpolME}, we illustrate our two procedures -- linear interpolation between neighboring Ioffe times or a polynomial fit to the full Ioffe time dependence (five points corresponding to our five nucleon boosts).
For the latter, we find that quadratic or cubic fits are enough to provide good description of the data (in terms of $\chi^2/{\rm dof}$ that is $\mathcal{O}(1)$ or smaller).
We find an overall consistency between different ways of interpolating, with slight tensions between quadratic and cubic polynomials in certain regions of $\nu$.
Below, we also compare these interpolations at the level of the final PDFs, but we take the linear interpolation between neighboring Ioffe times as our preferred method, as it is more conservative and always agrees with both polynomial orders in the fitting method.
Note that choosing a polynomial order for this interpolation introduces a mild model assumption and hence, a difference with respect to other polynomial orders should be considered as a systematic uncertainty.
Moreover, it is a priori not clear whether deviations from the fitted curve, observed for some Ioffe times (e.g.\ for the real part of reduced ITDs at $\nu\approx0.26$ and $(z/a)^2=1$) are mere statistical fluctuations or represent a physical effect and the linear interpolation chooses to keep such deviations and not smoothen them out.
However, given the general consistency between all methods, any of them appears to be acceptable at this level of precision.

\subsection{Evolved and matched ITDs}
The $1/z$ scale-dependence of the ITDs at a fixed Ioffe time is accounted for by the evolution equation (\ref{eq:evol}).
We choose to evolve all matrix elements to the scale corresponding to $\mu=2$ GeV, which will become our final renormalization scale after scheme conversion to the $\MSb$ scheme.
Inspection of the logarithm in Eq.~(\ref{eq:evol}) reveals that the scale $\mu=2$ GeV corresponds to ITDs being evolved to $z=2e^{-\gamma_E-1/2}/\mu\approx0.72a$, i.e.\ around 0.067 fm at our lattice spacing ($1/z\approx2.9$ GeV).
For the strong coupling constant, we take the 1-loop value at $\mu=2$ GeV, $\alpha_s/\pi\approx0.129$.
Below, we also investigate the dependence on the choice of this value by comparing the results with the ones obtained from $\alpha_s/\pi=0.1$, the latter taken in the quenched study of Ref.~\cite{Orginos:2017kos} and close to the value of Ref.~\cite{Joo:2019jct}, $\alpha_s/\pi\approx0.096$, corresponding to a higher-loop coupling used in phenomenology.

\begin{figure}[h!]
\begin{center}
    \includegraphics[width=0.48\textwidth]{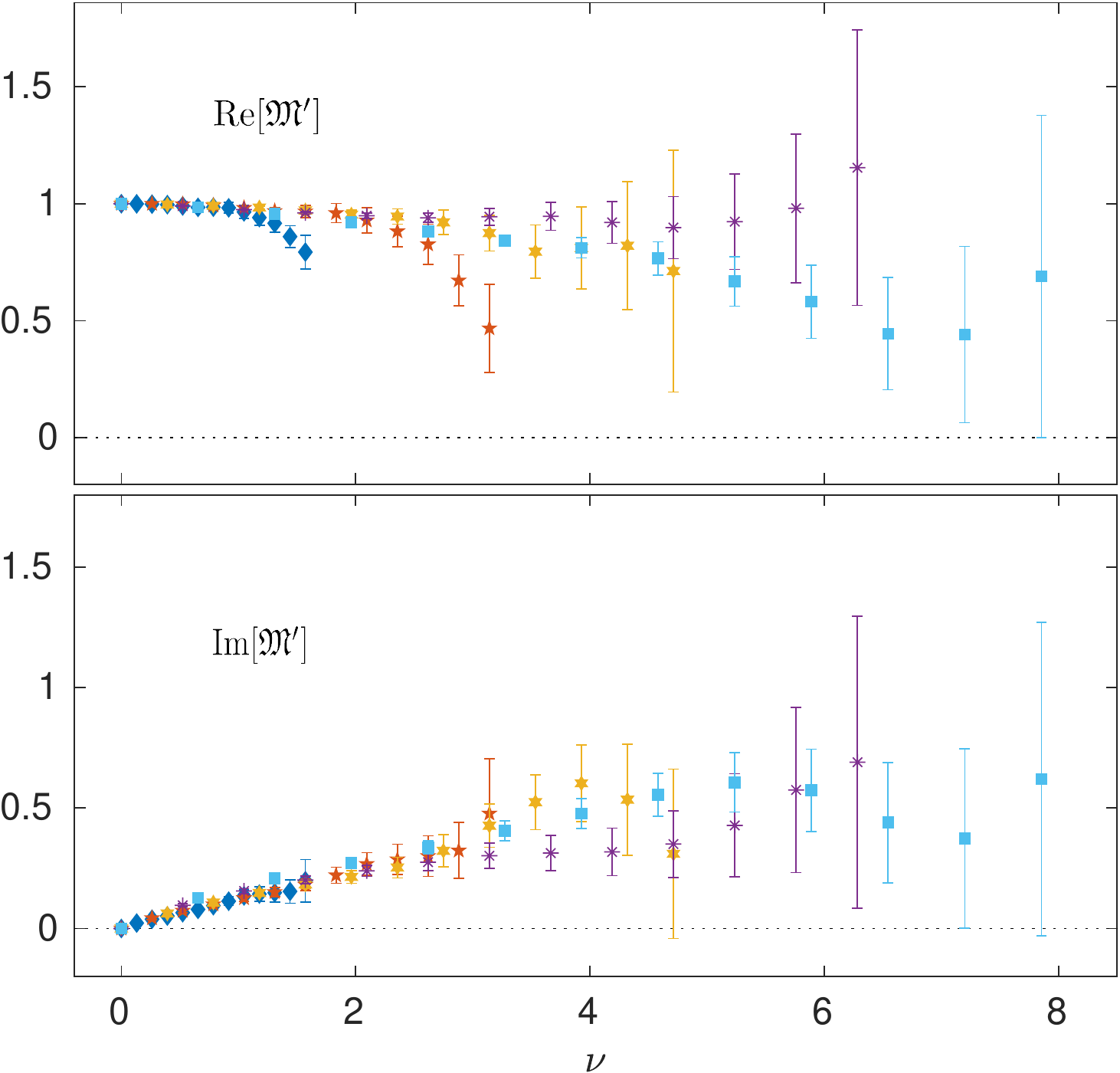}
\end{center}
\vspace*{-0.5cm} 
\caption{Real (top) and imaginary (bottom) part of the evolved matrix elements ($\mathfrak{M}'(\nu,z^2,\mu^2)$ at fixed $P_3$) for the unpolarized PDF. The scale after evolution is $1/z=\mu e^{\gamma_E+1/2}/2\approx2.9$ GeV and corresponds to the $\MSb$ scale $\mu=2$ GeV. Symbols are the same as used in Fig.~\ref{fig:bare}.}
\label{fig:evolved}
\end{figure}

The evolved ITDs are slightly closer to a universal curve (see Fig.~\ref{fig:evolved}), which is manifested by agreement between ITDs obtained from different combinations of $(P_3,z)$ at the same $\nu$ up to $z/a=10$ in the real part and for all values of $z/a$ in the imaginary part.

\begin{figure}[h!]
\begin{center}
    \includegraphics[width=0.48\textwidth]{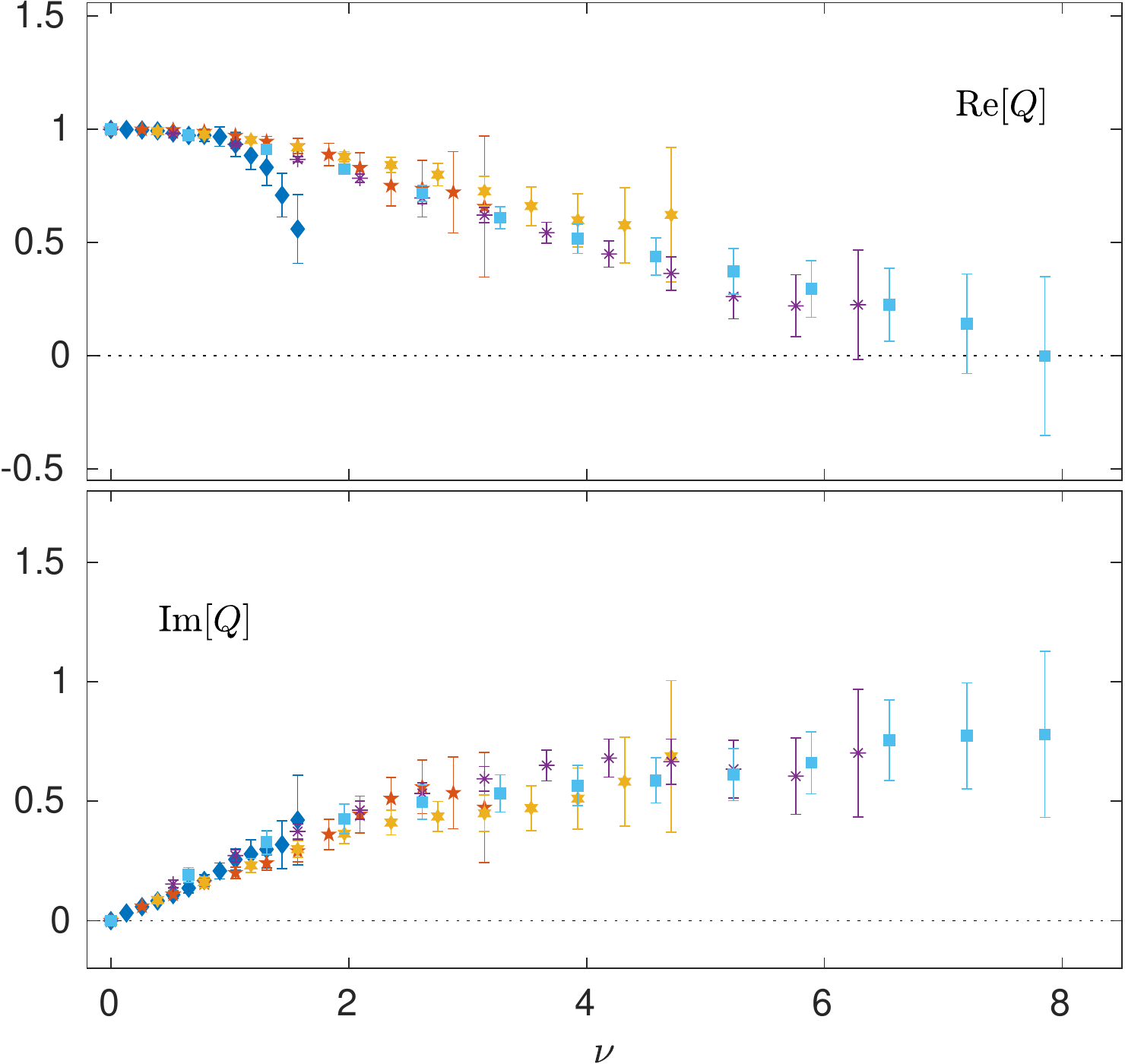}
\end{center}
\vspace*{-0.5cm} 
\caption{Real (top) and imaginary (bottom) part of the matched $\MSb(\mu=2$ GeV) matrix elements ($Q(\nu,z^2,\mu^2)$ at fixed $P_3$) for the unpolarized PDF. Symbols are the same as used in Fig.~\ref{fig:bare}.}
\label{fig:matched}
\end{figure}

The final step of the procedure to arrive at light-cone ITDs is to perform the scheme conversion according to Eq.~(\ref{eq:match}).
The outcome is shown in Fig.~\ref{fig:matched}.
The agreement between data at a given Ioffe time coming from different momenta and lengths of the Wilson line holds, similarly as for evolved ITDs, up to around 9-10 lattice spacings in the real part and for all values of $z$ in the imaginary part.

\begin{figure}[h!]
\begin{center}
    \includegraphics[width=0.48\textwidth]{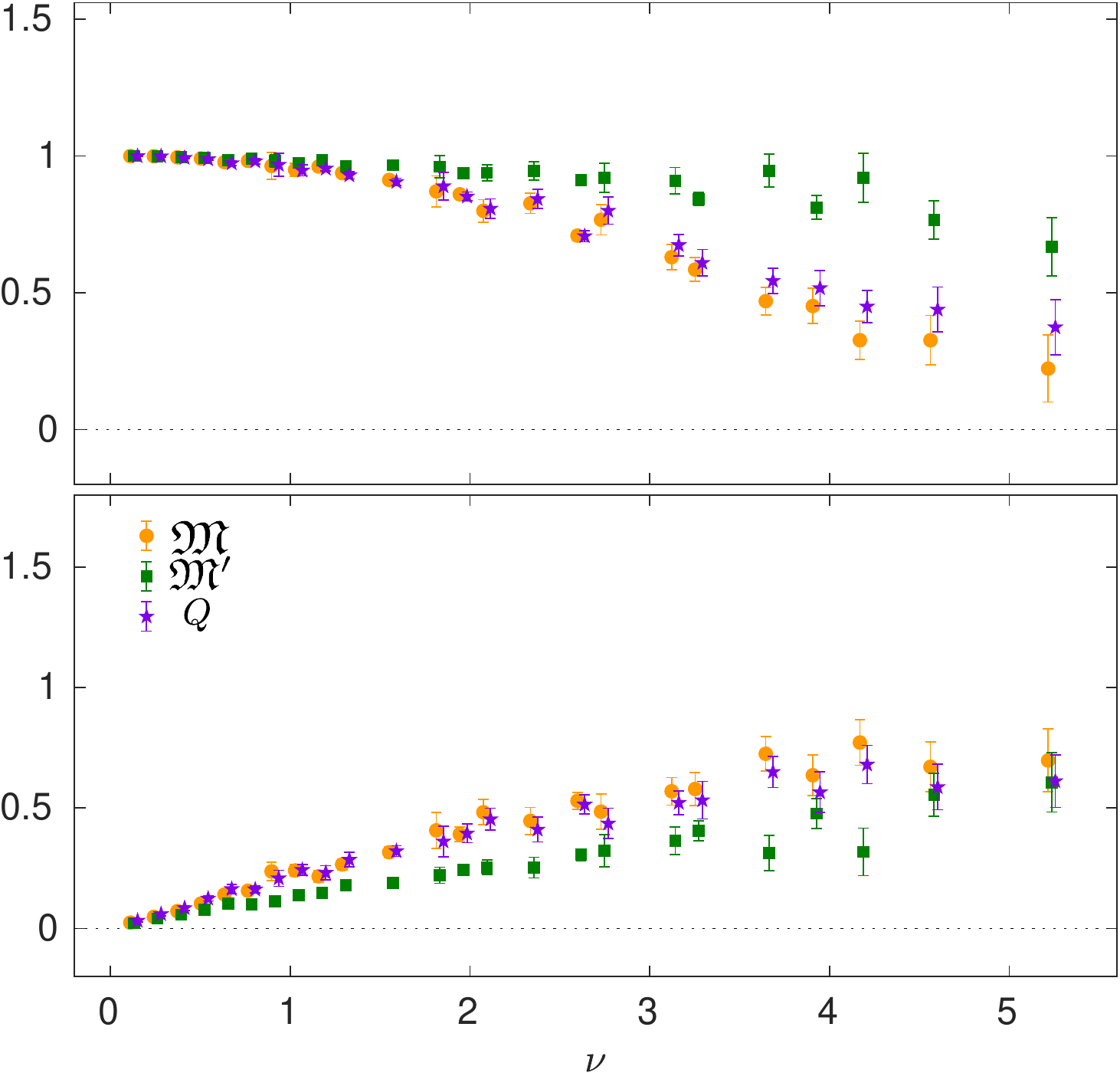}
\end{center}
\vspace*{-0.5cm} 
\caption{Real (top) and imaginary (bottom) part of the reduced ($\mathfrak{M}(\nu,z^2)$; orange circles), evolved ($\mathfrak{M}'(\nu,\mu^2)$; green squares) and matched ($Q(\nu,\mu^2)$; purple stars) Ioffe time distributions. The matrix elements corresponding to the same Ioffe time $\nu$ coming from different combinations of $(P_3,z)$ were averaged, keeping only the ones at $z/a\leq8$.}
\label{fig:ME_comparison}
\end{figure}

The effects of evolution and scheme conversion are summarized in Fig.~\ref{fig:ME_comparison}, where we averaged ITDs corresponding to the same Ioffe time $\nu$, but originating from different combinations of $(P_3,z)$.
To avoid contamination from ITDs that are off from a universal curve, we restricted the average to cases with $z/a\leq8$.
The effects of evolution and scheme conversion are opposite to each other and, interestingly, approximately equal in magnitude.
Thus, the final matched ITDs turn out to be compatible, within our statistical precision, with original reduced matrix elements.
It implies that the one-loop matching procedure at the level of ITDs is a small effect, which raises hope that higher-order matching effects are even smaller.
Nevertheless, obviously, a two-loop computation is still desired to check explicitly this statement.
It is also worth to contrast the one-loop matching for the case of pseudo-PDFs with the one for quasi-PDFs.
In the latter case, as e.g.\ in Ref.~\cite{Alexandrou:2019lfo} that used the same bare matrix elements as the present study, the one-loop matching effects are considerably larger.
The matching for quasi-PDFs is performed in $x$-space and the difference between a quasi-PDF and a matched PDF are above 100\% in many regions of $x$.
Hence, it is plausible that the matching in the pseudo-distribution approach, at the level of ITDs (in $\nu$-space), is more controlled, i.e.\ less subject to truncation effects.

\begin{figure}[h!]
\begin{center}
    \includegraphics[width=0.48\textwidth]{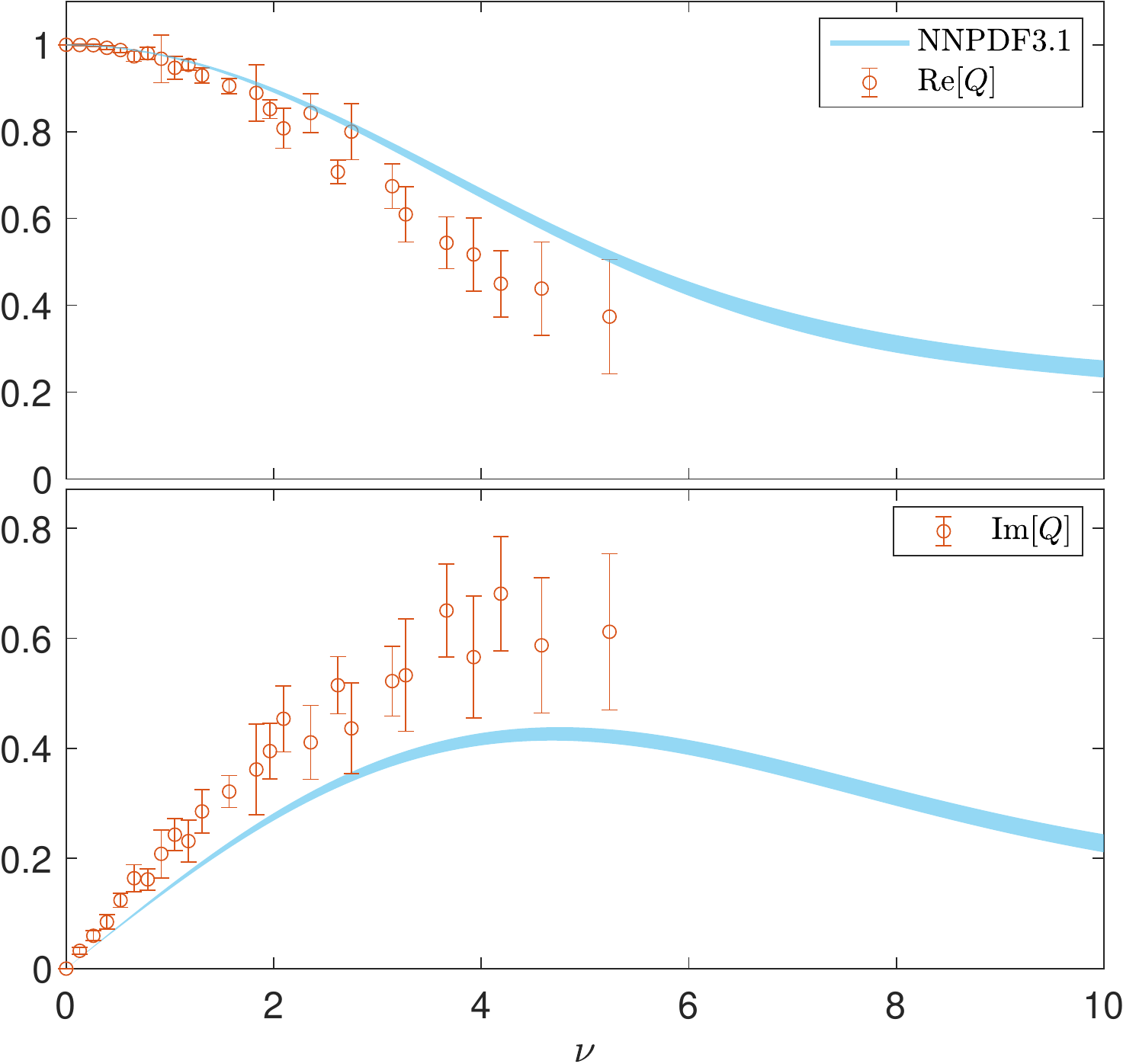}
\end{center}
\vspace*{-0.5cm} 
\caption{Real (top) and imaginary (bottom) part of the matched Ioffe time distributions (open circles; $\zmax/a=8$, $\alpha_s/\pi\approx0.129$, $\MSb$ scheme at $\mu=2$ GeV) compared to NNPDF distributions \cite{Ball:2017nwa} at the same renormalization scale, inverse-Fourier-transformed to $\nu$-space. }
\label{fig:NNPDF_ITDs}
\end{figure}

It is interesting to compare to phenomenological distributions already at the level of ITDs, which can be done by Fourier-transforming these distributions back to $\nu$-space.
The inverse Fourier transform, an integral of a continuous $x$-dependent distribution, is not subject to an inverse problem and hence, in principle, one can conclude about the agreement of lattice and phenomenology without inverse problem issues involved in reconstructing an $x$-dependent distribution from a finite set of $\nu$-truncated lattice ITDs.
In Fig.~\ref{fig:NNPDF_ITDs}, we compare our matched ITDs ($\zmax/a=8$, $\alpha_s/\pi\approx0.129$, $\MSb$ scheme at $\mu=2$ GeV) to inverse-Fourier-transformed NNPDF distributions \cite{Ball:2017nwa}.
The matched ITDs are shown with their statistical errors, as well as with estimated systematic uncertainties, to be defined and discussed below in Sec.~\ref{sec:final}.
We conclude good agreement of the lattice-extracted ITDs and the phenomenological curves, with a slight tendency of the lattice ITDs to decay a bit faster than NNPDF for the real part and the opposite tendency in the imaginary part.
Nevertheless, the agreement is at the level of 1-$\sigma$ to 2-$\sigma$ for ITDs at most Ioffe times.
Overall, this agreement between lattice and phenomenological ITDs is reasonable, giving good prospects for the reconstruction of $x$-dependent distributions.
However, there are indications that various systematic effects may be sizable.
Moreover, it is clear from the slow decay of NNPDF ITDs that a robust and unambiguous reconstruction of the $x$-dependence may only be achieved when the lattice calculations can get to significantly larger Ioffe times, particularly for distributions involving the imaginary part.

\begin{figure*}[t!]
\begin{center}
    \includegraphics[width=0.9\textwidth]{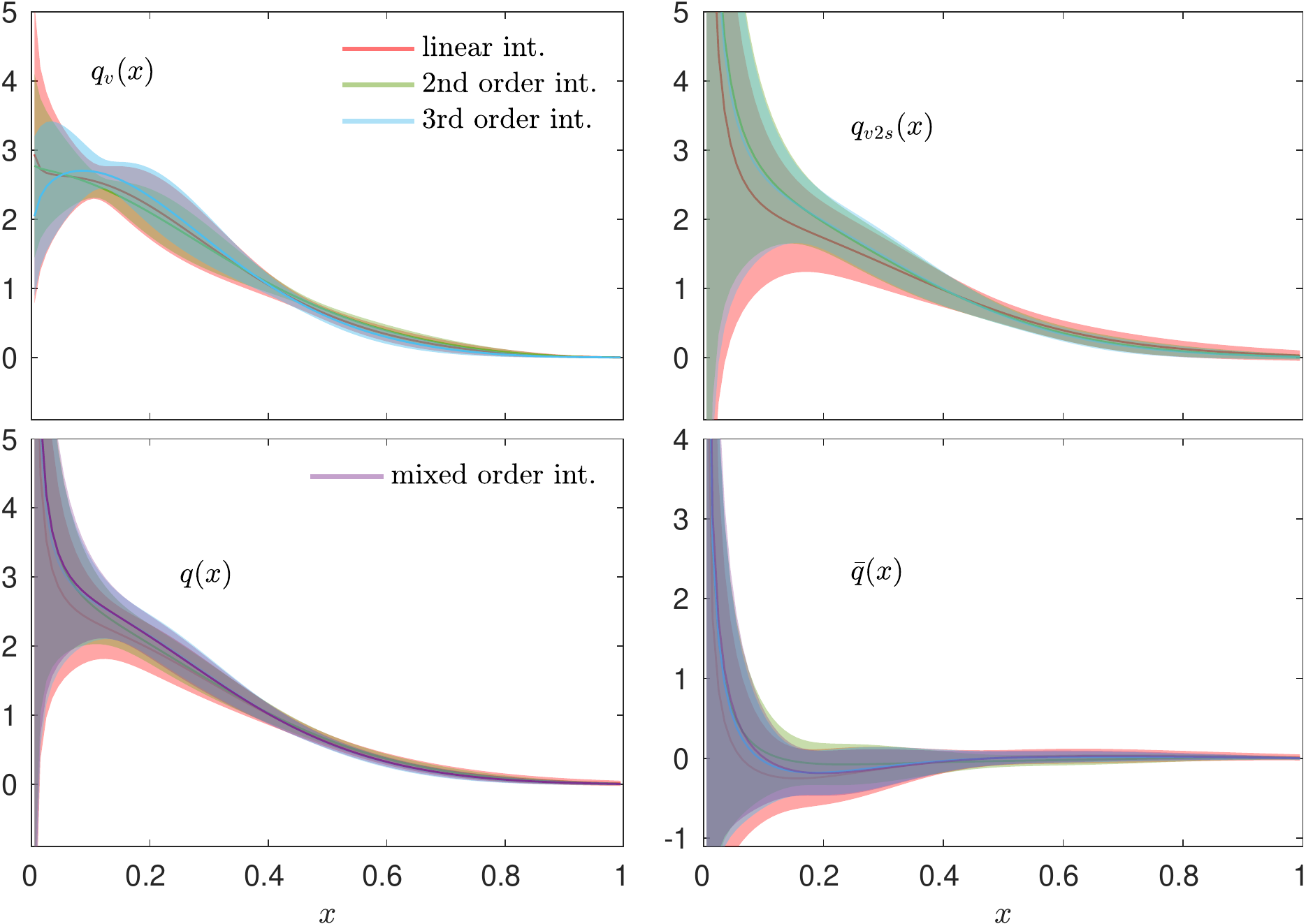}
\end{center}
\vspace*{-0.5cm} 
\caption{PDFs obtained with different interpolation methods to access reduced ITDs at continuous Ioffe times and fixed $z^2$: linear interpolation between ITDs at neighboring Ioffe times (red), second/third order polynomial interpolations from fits to the full Ioffe time dependence (green/blue) and, where applicable, mixed second/third order for the imaginary/real part of ITDs (purple).
Light-cone distributions $q_v$ (top left), $q_{v2s}$ (top right), $q$ (bottom left), $\bar{q}$ (bottom right). The ITDs are fitted up to $\numax\approx5.2$ ($\zmax/a=8$), $\alpha_s/\pi\approx0.129$, $\MSb$ scheme at $\mu=2$ GeV.}
\label{fig:interpol}
\end{figure*}

\begin{figure*}[t!]
\begin{center}
    \includegraphics[width=0.9\textwidth]{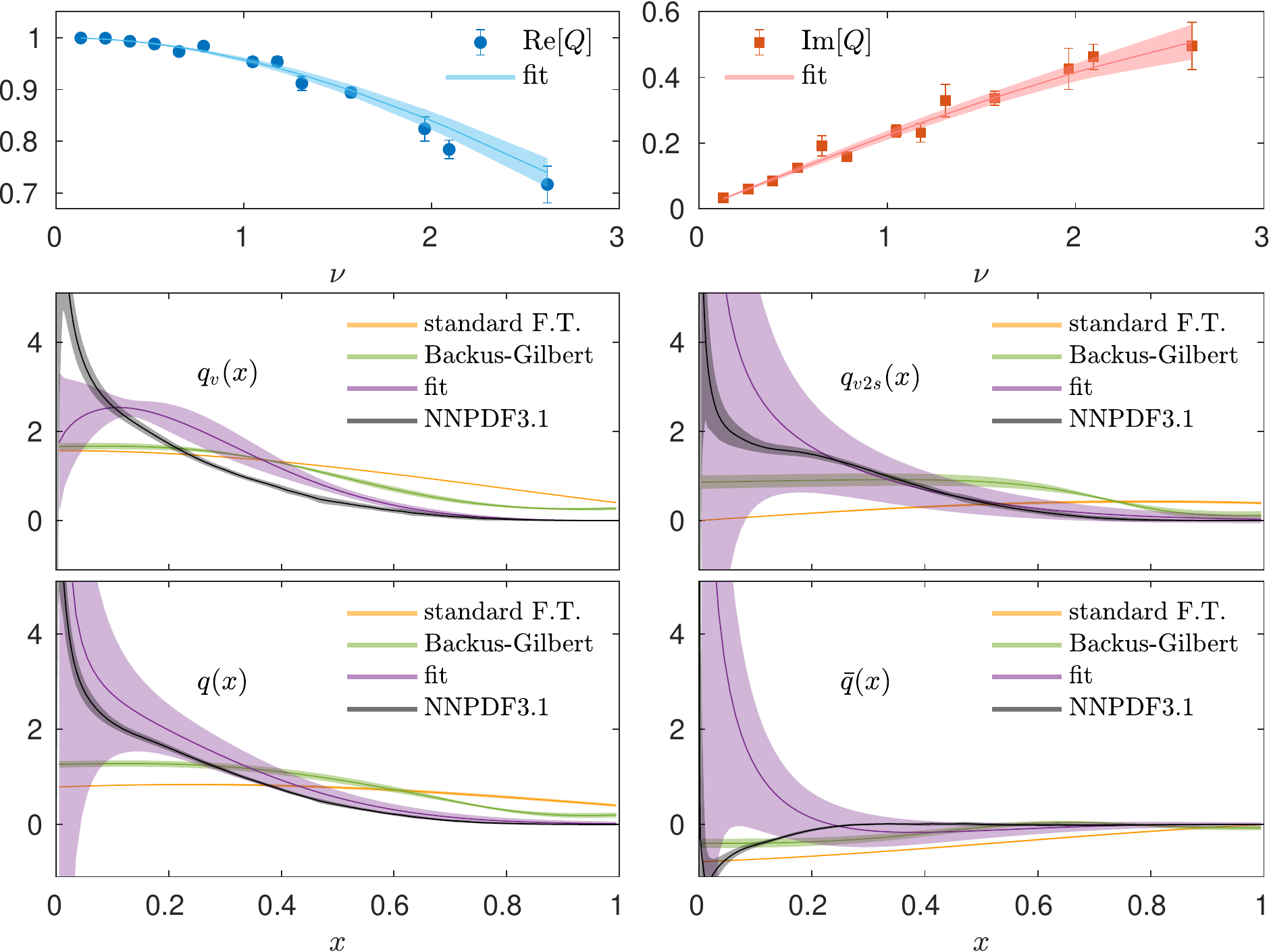}
\end{center}
\vspace*{-0.5cm} 
\caption{Top row, left/right panels: real/imaginary part of matched ITDs (blue circles / red squares) and fitted ITDs (blue/red band).
Middle and bottom row: light-cone distributions $q_v$ (middle left), $q_{v2s}$ (middle right), $q$ (bottom left), $\bar{q}$ (bottom right) from 3 reconstruction methods: naive Fourier transform (orange), Backus-Gilbert (green), fitting ansatz for the distribution (purple). Shown are also NNPDF phenomenological distributions (grey) \cite{Ball:2017nwa}. The ITDs are fitted up to $\numax\approx2.6$ ($\zmax/a=4$), $\alpha_s/\pi\approx0.129$, $\MSb$ scheme at $\mu=2$ GeV.}
\label{fig:zmax4}
\end{figure*}

\subsection{Light-cone PDFs}
We now present results for the unpolarized PDFs obtained from matched ITDs discussed in the previous subsection.

Reconstruction of a PDF from ITDs requires, in principle, the knowledge of the full Ioffe time dependence of the ITDs, from $\nu=0$ to $\nu=\infty$.
Obviously, with numerical calculations of ITDs on the lattice, the upper limit, $\numax$, is necessarily finite.
It is desirable to take $\numax$ as large as possible, ideally to observe that ITDs have decayed to zero.
However, as Fig.~\ref{fig:matched} suggests, this is difficult with the currently attained nucleon boosts.
The real part of matched ITDs approaches zero at $\nu\approx7-8$, but these Ioffe times are obtained with Wilson line lengths of order 1 fm at 
the largest boost, which corresponds to very low energy scales at which the matching procedure is likely to fail.
For the imaginary part of ITDs, non-zero values are observed even at $\nu=8$.
It is, thus, clear that a robust extraction of the full $x$-dependence requires achieving even larger Ioffe times, especially for the imaginary part that yields the distribution $q_{v2s}$.
Since the length of the Wilson line is limited by the reliability of the matching procedure and HTE of $\mathcal{O}(z^2\LambdaQCD^2)$, larger Ioffe times need to be reached at larger nucleon boosts.
This is, however, difficult for the lattice, which is due to the decaying signal-to-noise ratio with increasing nucleon momentum, as discussed in Ref.~\cite{Alexandrou:2019lfo}.

\begin{figure*}[t!]
\begin{center}
    \includegraphics[width=0.9\textwidth]{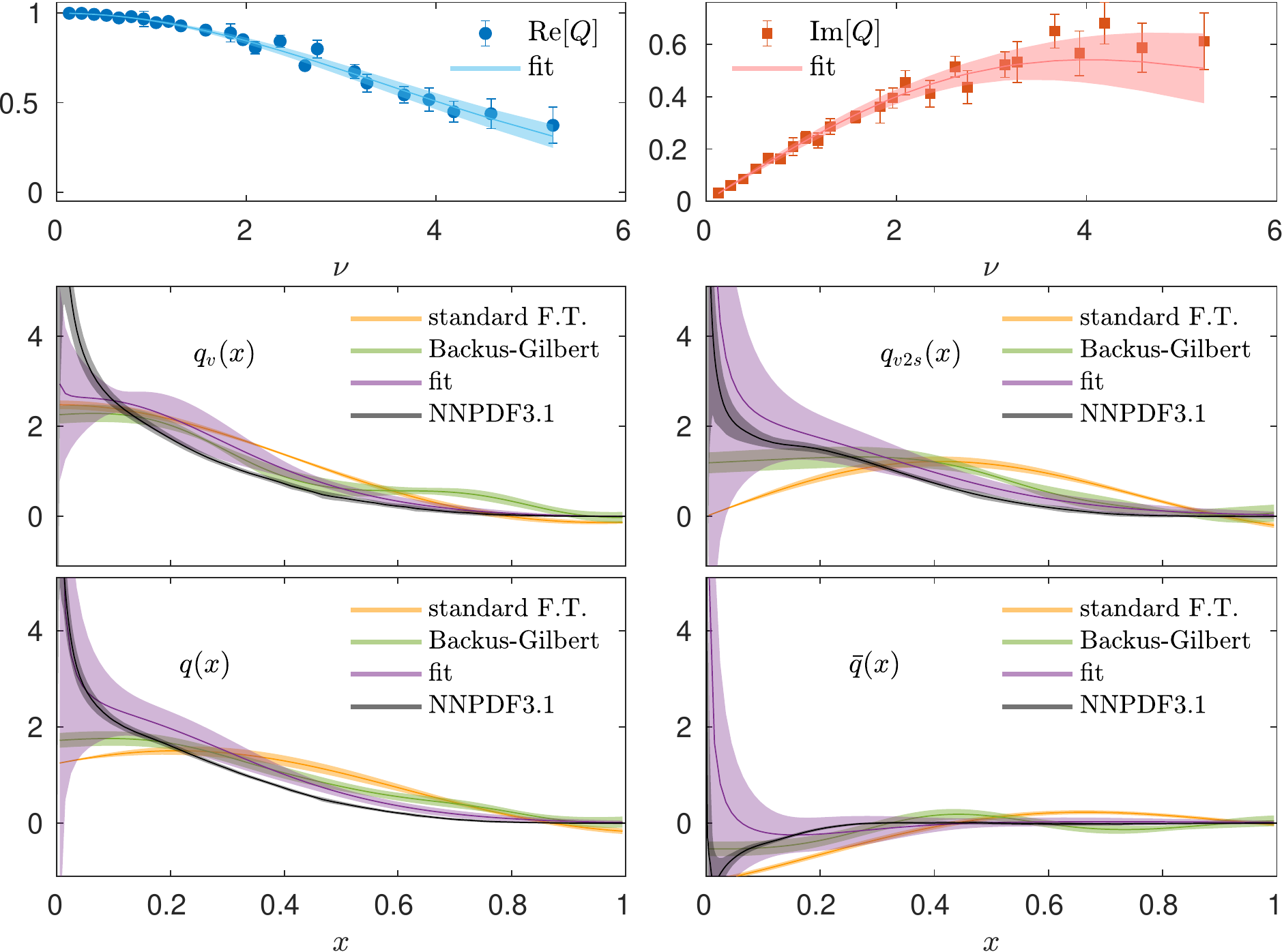}
\end{center}
\vspace*{-0.5cm} 
\caption{Same as Fig.~\ref{fig:zmax4}, but the range of Ioffe times extended to $\numax\approx5.2$ ($\zmax/a=8$).}
\label{fig:zmax8}
\end{figure*}

\begin{figure*}[t!]
\begin{center}
    \includegraphics[width=0.9\textwidth]{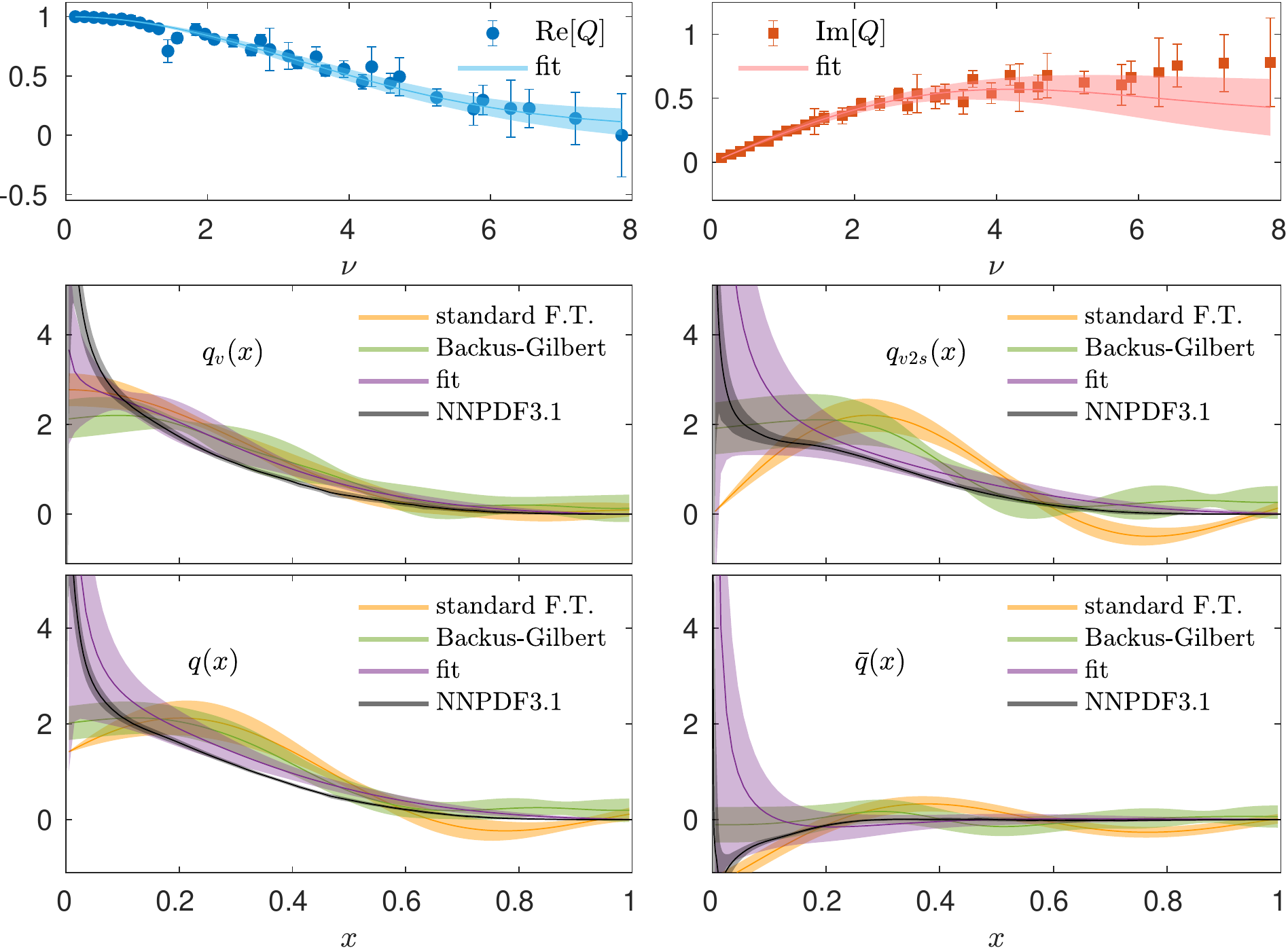}
\end{center}
\vspace*{-0.5cm} 
\caption{Same as Fig.~\ref{fig:zmax4}, but the range of Ioffe times extended to $\numax\approx7.9$ ($\zmax/a=12$).}
\label{fig:zmax12}
\end{figure*}

\begin{figure*}[p!]
   \centering
   \setlength{\unitlength}{0.1\textwidth}
   \begin{picture}(10,10)
     \put(3.1,5.6){\includegraphics[width=0.66\textwidth]{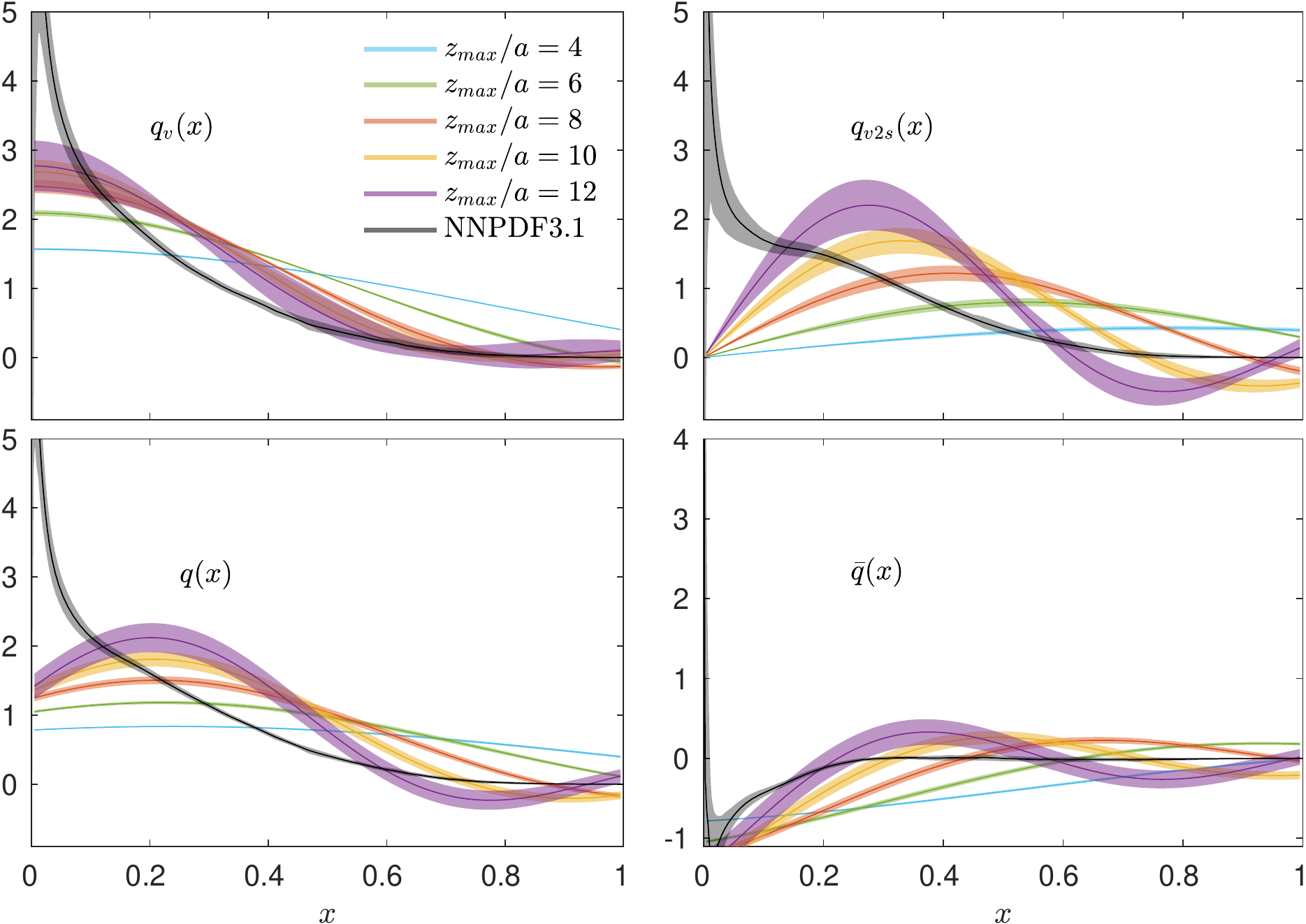}}
     \put(0.4,9.8){\huge Naive FT}
     \put(3.1,1){\includegraphics[width=0.66\textwidth]{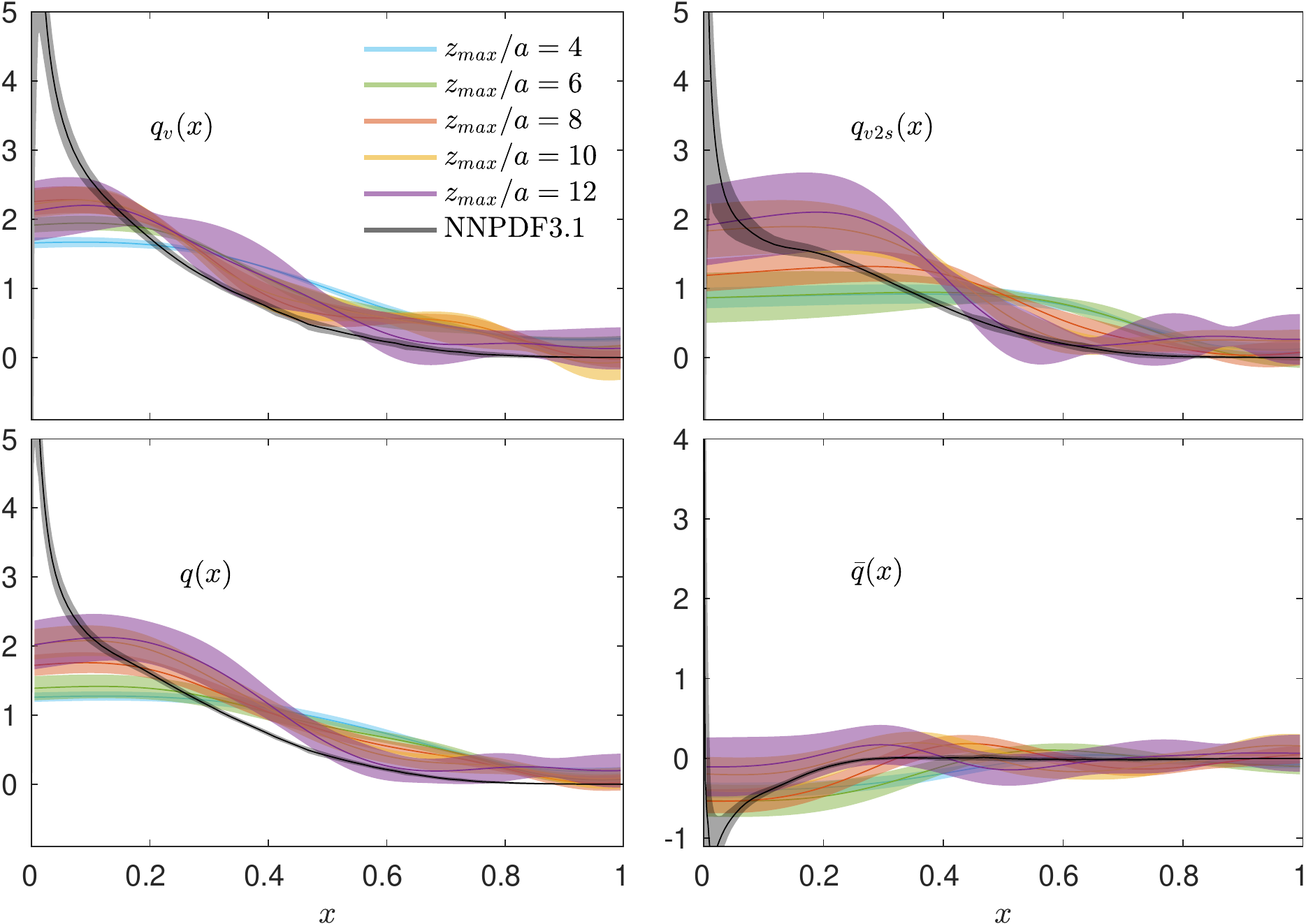}}
     \put(0.4,5.3){\huge Backus-Gilbert}
     \put(3.1,-3.6){\includegraphics[width=0.66\textwidth]{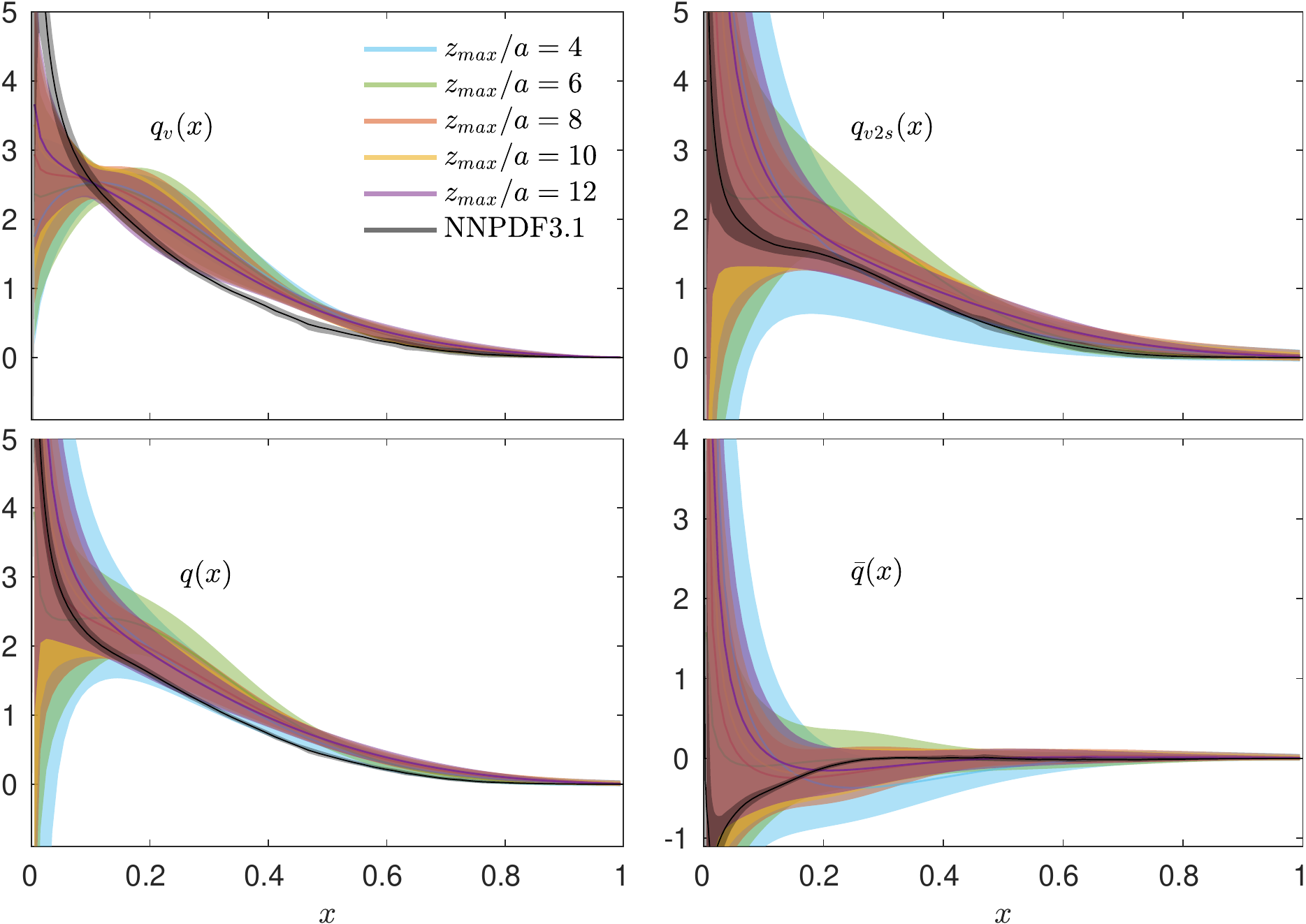}}
     \put(0.4,0.7){\huge fits}
   \end{picture}
\vspace*{1.8cm}   
\caption{The dependence of the\\ reconstructed distributions on the \\ range of available Ioffe times,\\ proxied by $\zmax/a$.\\
The upper 4 plots are for $q_v$, $q_{v2s}$,\\
$q$, $\bar{q}$ with the naive Fourier transform,\\ the middle 4 for the Backus-Gilbert \\
method and the bottom 4 for \\ the fitting ansatz reconstruction.}
\label{fig:zmax}
\end{figure*}

\begin{figure*}[t!]
\begin{center}
    \includegraphics[width=0.9\textwidth]{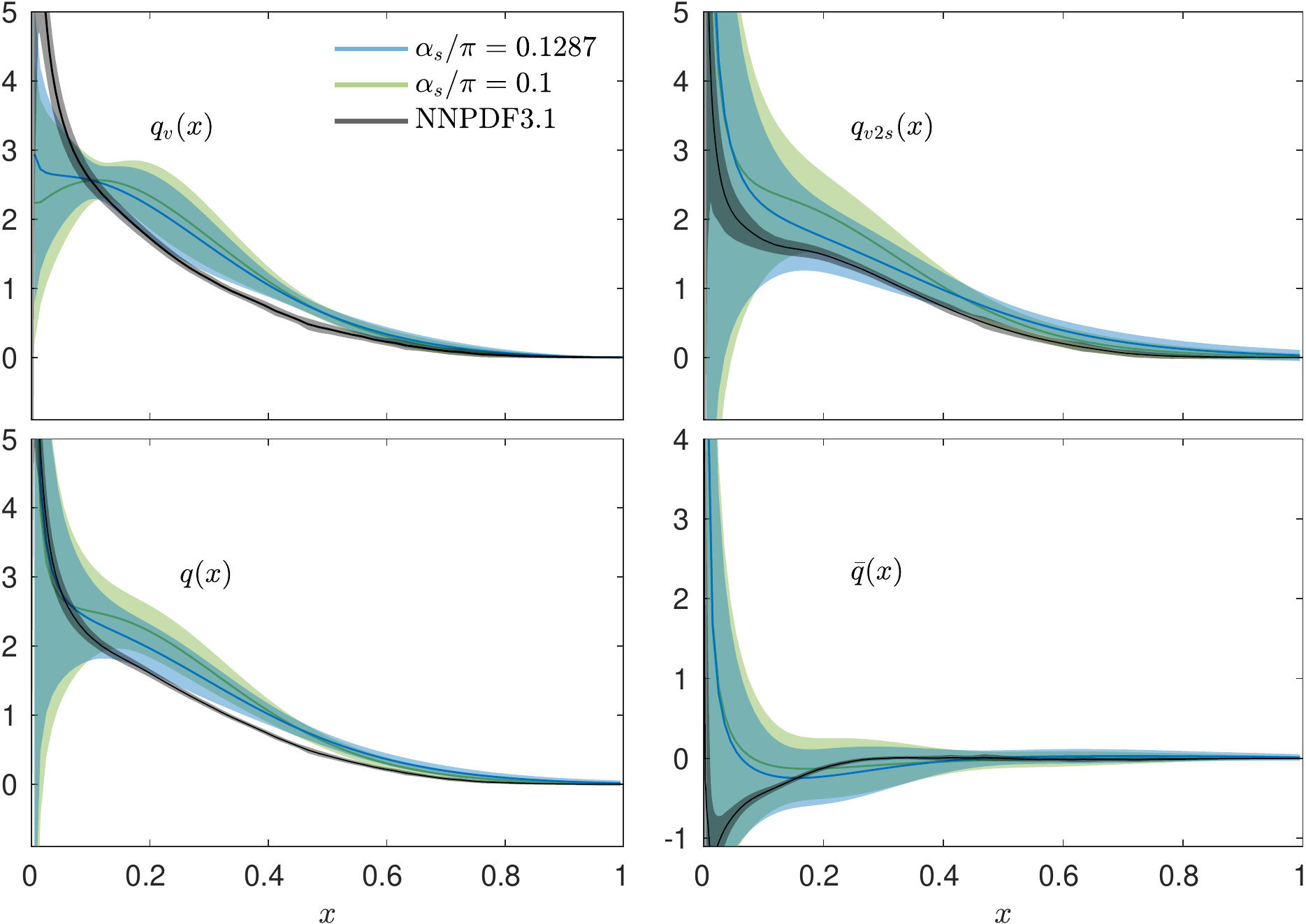}
\end{center}
\vspace*{-0.5cm} 
\caption{The dependence of the fitting ansatz reconstructed distributions (upper left: $q_v$, upper right: $q_{v2s}$, lower left: $q$, lower right: $\bar{q}$) on the value of the strong coupling constant, $\alpha_s/\pi$. The range of Ioffe times in the reconstruction is up to $\numax\approx5.2$ ($\zmax/a=8$).}
\label{fig:alpha}
\end{figure*}

Meanwhile, we are in position to investigate what the currently available range of Ioffe times implies for the $x$-dependence.
The key parameter to decide in reconstructing PDFs is the maximum Ioffe time, $\numax$.
Below, we provide reconstructed PDFs for various choices of $\numax$, ranging from around 2.6 to 7.9.
The former corresponds to taking ITDs obtained from matrix elements with insertions of the operator with Wilson line length up to $\zmax/a=4$ lattice units (around 0.37 fm) and the latter to 12 lattice units (1.12 fm).
It is unclear, a priori, which value of $\numax$ ensures reliable matching and good control over HTE.
However, given the statistical uncertainties of our data, we adopt a criterion that safe values of $\numax$ are those for which the matched ITDs obtained from different combinations $(P_3,z)$ corresponding to the same $\nu$ are consistent with each other.
This criterion leads to maximum $z$ of order 8-9 lattice units (approx.\ 0.8 fm), as we have discussed in the context of Fig.~\ref{fig:matched}.
In this way, the reached Ioffe times are of order 5-6.
At these values of $\nu$, the real part of matched ITDs is already close to 0, thus giving good hope for the reconstruction of the valence distribution.
In turn, the imaginary part of ITDs is still rather far away from zero, which is expected to bring significant uncertainties particularly into $q_v+2\bar{q}$.

Before we investigate the $\zmax$-dependence, we compare PDFs at a selected value of $\zmax=8a$ obtained with different interpolation prescriptions to access reduced ITDs at continuous Ioffe times.
This is illustrated in Fig.~\ref{fig:interpol}, where we show the considered four types of PDFs from linear interpolation between neighboring Ioffe times as well as from polynomial interpolations of second and third order.
For the cases of PDFs that mix real and imaginary parts of ITDs ($q$ and $q_s$), we also show a mixed-order interpolation (second/third order for imaginary/real part), motivated by the fact that second order polynomials are enough to obtain good fits to the imaginary part at all values of $z^2$.
As anticipated from comparisons at the level of ITDs, all interpolation methods give consistent PDFs.
The approach of the linear interpolation between ITDs at neighboring Ioffe times leads to largest errors and we follow it below as the most conservative choice.

In Figs.~\ref{fig:zmax4}, \ref{fig:zmax8} and \ref{fig:zmax12}, we show the reconstructed PDFs with $\zmax/a=4$, 8 and 12, respectively ($\numax\approx2.6$, $5.2$ and $7.9$).
The matrix elements used in the reconstruction are shown in the upper row of each figure.
It is clear that if $\zmax/a=4$ (Fig.~\ref{fig:zmax4}), the data cover an insufficient range of Ioffe times and thus, the naive Fourier transform, as well as the BG method are simply missing the data and lead to unrealistically looking distributions.
However, interestingly, the PDF fitting ansatz approach provides significantly better distributions.
The data at small Ioffe times are very precise and guide the fits, leading to precise results at large $x$, in full agreement with the phenomenological curves of NNPDF \cite{Ball:2017nwa} for all distributions.
The fitted matrix elements, $Q_f(\nu,\mu^2)$, are depicted as bands in the upper row of Fig.~\ref{fig:zmax4} and the fits provide good description of the data ($\chi^2/{\rm dof}\approx1.0$ ($0.5$) for the fit of the real (imaginary) part).
The missing data at large $\nu$ manifest themselves in the increasing uncertainty of the reconstructed PDFs at low $x$, in particular for the $q_{v2s}$ distribution coming from the imaginary part of matched ITDs.
The latter uncertainty propagates itself also to the full distribution $q=q_v+\bar{q}$ and to the sea distribution $\bar{q}$, obtained from linear combinations of $q_v$ and $q_{v2s}$.
We note that the full distribution $q$ is consistent with NNPDF for all values of $x$, which, however, holds for $x\lesssim0.2$ within rather large uncertainties and is partially accidental -- the valence distribution is significantly above the curve from phenomenological fits for a wide range of $x$, but this difference is compensated in $q$ by the perfect agreement of $q_{v2s}$ in this range.

We want to see now how robust are the PDFs obtained with $\numax\approx2.6$ when increasing the range of Ioffe time used in the fits and whether the other two methods of PDF reconstruction can lead to conclusive distributions.
In Fig.~\ref{fig:zmax8}, we show the case of $\zmax/a=8$, which leads to $\numax\approx5.2$.
This range of Ioffe times contains significantly more data points than the range up to $\numax\approx2.6$ and thus, distributions reconstructed using the naive Fourier transform and, particularly, the BG method start evincing qualitative features of phenomenological PDFs.
The additional data at larger $\nu$ entering the fits ($\chi^2/{\rm dof}\approx0.6$ ($0.4$) for the fit of the real (imaginary) part) provide more constraints for fitting parameters and their most significant effect is to decrease the PDFs uncertainty at $x\lesssim0.4-0.5$.
This originates from the reduced uncertainty in the fitted matrix elements, i.e.\  the smaller widths of bands in the upper row of Fig.~\ref{fig:zmax8} in the region $\nu\approx2.6-5.2$, now constrained by the actual lattice data and not simply guided by the low-$\nu$ behavior.
However, the bands overlap for the cases of $\zmax/a=4$ and $\zmax/a=8$ and thus, the resulting PDFs move only within the uncertainties of the former case.
This is encouraging, since it suggests relatively little dependence on $\numax$, with the effect of increasing the latter restricted predominantly to giving more precise access to lower $x$ values for the distributions.

Similar conclusions are drawn when further increasing the range of Ioffe times, to $\numax\approx7.9$ ($\zmax/a=12$).
For this case, shown in Fig.~\ref{fig:zmax12}, one needs to keep in mind that the contamination of the large-$z$ ITDs might be significant, as best seen in the deviation of the $z/a\gtrsim10$ lowest-momentum points from an universal curve (Fig.~\ref{fig:matched}).
The extended $\nu$-range has a similar effect on the distributions as the one when increasing $\numax$ from 2.6 to 5.2 -- PDFs extracted with naive Fourier transform and with the BG method are now qualitatively closer to the phenomenological distributions.
The fitting reconstruction again provides good description of the matched ITDs ($\chi^2/{\rm dof}\approx0.5 ($0.3) for the real (imaginary) part) and PDFs reconstructed by fitting are only slightly changed with respect to $\zmax/a=8$.
In fact, they are also compatible with the case of the shortest range of Ioffe time, $\zmax/a=4$, within uncertainties.
The larger range in Ioffe times again decreases the uncertainty in the low- and intermediate-$x$ regions.
For instance, at $x=0.2$, the error in $q_{v2s}$ is approximately twice smaller with $\numax\approx7.9$ as compared to $\numax\approx2.6$.
An important, if somewhat obvious, effect of enhancing the $\nu$-range is that the inverse problem in the distribution reconstruction becomes less ill-defined.
This manifests itself in the gradual convergence of all 3 reconstruction methods.
At $\zmax/a=4$, the amount of information in the lattice data is scarce and the rather mild additional assumption implicit in the BG method helps only little.
The physically-motivated assumption provided by the fitting ansatz is, in turn, enough to reconstruct particularly the large-$x$ part of the distributions.
When increasing $\zmax$, there is more information from the lattice data that shapes the functional form of the distributions and the mild assumption of the BG method is enough to obtain PDFs consistent with the ones from the fitting ansatz.
This holds in the full $x$-range, for all considered distributions.
The stronger assumption contained in the fitting ansatz is, in turn, verified by adding larger-$\nu$ data, proving that the estimates of large-$x$ PDFs are robust.
The access to low $x\lesssim0.2$ is still limited even with $\numax\approx7.9$ and would require data at yet larger Ioffe times.

The dependence of the results on the Ioffe time range for the 3 reconstruction methods is summarized in Fig.~\ref{fig:zmax}, where we show distributions obtained with $\zmax/a=4,\,6,\,8,\,10,\,12$.
It clearly demonstrates that the naive Fourier transform is not a plausible method to reconstruct the $x$-dependence.
All naively reconstructed distributions are unrobust against changing the $\nu$-range and, at best, certain qualitative agreement is observed with phenomenological PDFs when $\zmax$ is large.
The BG method does significantly better already with intermediate values of $\zmax$.
The qualitative features of the PDFs are reproduced, especially at larger $x$.
With large $\zmax$, there is even quantitative agreement with NNPDF for a wide range of $x$.
Comparing $q_v$ and $q_{v2s}$, the former is more robustly reconstructed due to the faster decay of the real part of matched ITDs in $\nu$ -- effectively, there is more ``missing information'' in the imaginary part.
This missing data is also manifested in the irregular behavior of the error, with PDFs at some $x$ having artificially suppressed or enhanced errors.
It is rather evident that the most robust way of reconstructing the PDFs is the fitting ansatz approach.
The numerically stronger assumption for regulating the inverse problem than the one in the BG method, but physically a well-motivated one, leads to a regular behavior of PDFs with respect to the range of Ioffe times.
Distributions obtained from all values of $\numax$ are compatible with one another and the parameter $\numax$ predominantly controls the uncertainties, in general decreasing the errors at lower values of $x$ when data at larger $\nu$ are included.
This is particularly visible for distributions involving the imaginary part of ITDs.
For the valence distribution, related to the real part of ITDs by a cosine Fourier transform, the data at large Ioffe times are suppressed -- the real part of ITDs decays more quickly in $\nu$ than the imaginary part and moreover, the cosine function gives more weight to small Ioffe times (apart from very small $x$).
As a consequence, the valence distribution is very robust against $\numax$, with reduction of uncertainties with increasing $\numax$ visible only for $x\lesssim0.1$.
In turn, all other distributions receive contributions from the imaginary part of ITDs, weighed by $\sin(x\nu)$.
When extending $\numax$ from 2.6 to 5.2 and 7.9, these contributions are numerically potentially more important compared to the case of $q_v$, since both the weight is numerically larger for a wide range of $x$ and the imaginary part of ITDs is not suppressed in this regime of Ioffe times.
However, we note that after a linear rise of ${\rm Im}\,Q(\nu)$ until $\nu\approx3$, these ITDs are approximately constant for larger Ioffe times and thus, their overall contribution is suppressed by the periodicity of the sine weight.
Finally, we observe that the sensitivity to $\numax$ for distributions related to the imaginary part is only slightly larger than the one of $q_v$.

\begin{figure*}[t!]
\begin{center}
    \includegraphics[width=0.9\textwidth]{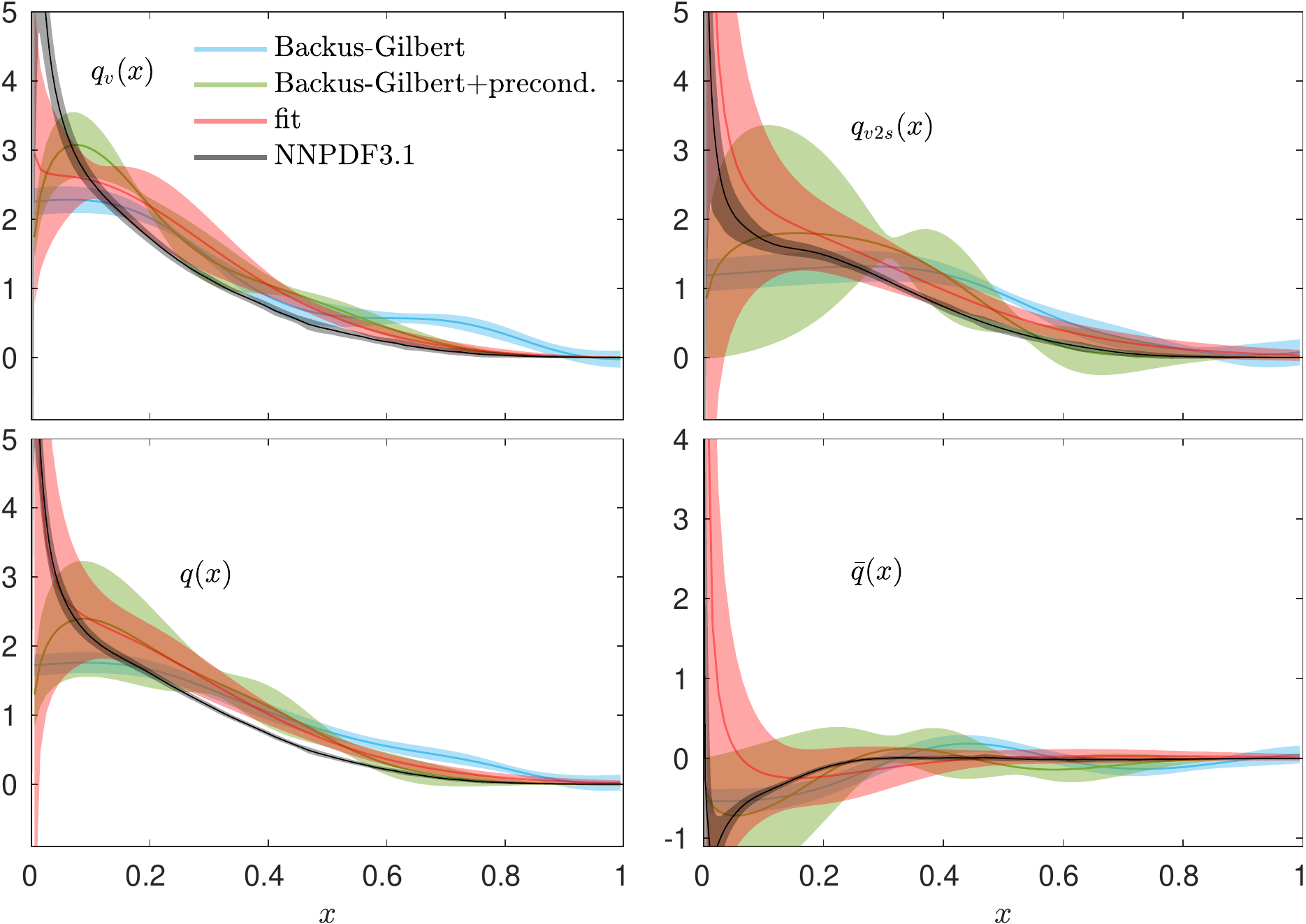}
\end{center}
\vspace*{-0.5cm} 
\caption{Comparison of the distributions (upper left: $q_v$, upper right: $q_{v2s}$, lower left: $q$, lower right: $\bar{q}$) reconstructed with the fitting ansatz approach (red) and the Backus-Gilbert method with (green) and without (blue) preconditioning. The range of Ioffe times in the reconstruction is up to $\numax\approx5.2$ ($\zmax/a=8$), $\alpha_s/\pi\approx0.129$.}
\label{fig:precond}
\end{figure*}

As a further check of systematics, we vary the strong coupling constant in the one-loop evolution and matching of ITDs.
In Fig.~\ref{fig:alpha}, we compare our choice of this coupling, $\alpha_s/\pi\approx0.129$ at the $\MSb$ scale of 2 GeV, with $\alpha_s/\pi=0.1$, equal to or close to the value used in other studies \cite{Orginos:2017kos,Joo:2019jct}.
Even though the change of $\alpha_s$ is rather large, the final distributions are not heavily affected.
The largest effect is observed for $x\approx0.1-0.3$ in $q_{v2s}$ and it gets propagated also to the full distribution $q$.
However, even this change is well within statistical uncertainties.
The robustness of the PDFs with respect to $\alpha_s$ is a consequence of the opposite sign of effects in the evolution and the scheme conversion.
While the evolved ITDs depend on $\alpha_s$ in a more pronounced way, the effect of the scheme conversion brings them back towards reduced matrix elements and thus, the dependence of the matched ITDs and of the light-cone PDFs on $\alpha_s$ is relatively mild.
It remains to be established whether this feature holds also at higher-loop orders.

The final question that we want to address in this subsection is how much the preconditioning affects the BG reconstruction.
Above, we observed that the BG results become increasingly consistent with the ones from the fitting reconstruction when $\numax$ is increased.
At $\numax\approx7.9$, both methods agree within uncertainties for the full $x$-range of all distributions, as illustrated in Fig.~\ref{fig:zmax12}.
Now, we test whether this agreement can be extended to lower $\numax$ when preconditioning the BG method with the function found in the fitting ansatz approach, i.e.\ the rescaling function $p(x)$ in Eq.~(\ref{eq:precond}) is of the form (\ref{eq:ansatz}) with parameters $a,b$ (for $q_v$) or $a,b,N$ (for $q_{v2s}$) taken to be the central values found in the fits.
We emphasize that this does not enforce such form of the distribution, but applies the BG criterion of maximal stability of the solution with respect to statistical variance of the data to the deviation of the distribution from the assumed one, instead of to the full distribution.
In this way, this tests the consistency of the BG assumption regulating the inverse problem with the assumption made in the fitting ansatz approach.
In Fig.~\ref{fig:precond}, the comparison of the distributions from the BG method with (green band) and without (blue band) preconditioning to the ones from the fits (red band) is given for $\numax\approx5.2$.
We note that preconditiong indeed increases the agreement between BG and fitting results and full consistency is observed between the two at all values of $x$.
However, the BG criterion for regulating the inverse problem is not completely equivalent to the fitting ansatz assumption, i.e.\ the reconstructed function $\tilde{q}(x)=q(x)/p(x)$ (see notation above Eq.~(\ref{eq:precond})) is not equal to 1.
The statistically most prominent effect of preconditioning is observed at large $x$ -- the preconditioned distributions evince now fully smooth behavior.
The agreement of PDFs from $\BGp$ with global fits is, however, slightly worse than for the fitting approach, particularly at $x\approx0.2$ in $q_v$.
As we noted already for the standard BG method in Fig.~\ref{fig:zmax}, one observes irregular behavior of the errors, which are artificially suppressed for some $x$ values and enhanced for others.
It can be interpreted that the statistical variance of our ITDs allows only for a very restricted value of the reconstructed PDF at some $x$, while at other values of $x$ significant variation is possible.
Naturally, such effects are linked to the missing data in Ioffe time.
We note that $\nu\in[0,5.2]$ is the minimum range of Ioffe times needed to observe consistency between $\BGp$ and fits.
This again points to the fact that $\numax\approx5.2$ is likely to be the optimal value for reconstructing the distributions, i.e.\ one giving the proper compromise for the Ioffe time range available with our maximal nucleon boost -- that should be large to provide enough information, but small enough to avoid contamination from HTE and unreliability of the matching procedure applied at too low scales.

\subsection{Final results with quantified systematic uncertainties}
\label{sec:final}
Having analyzed in detail three distribution reconstruction methods and the dependence of the results on the range of Ioffe times and on the value of the strong coupling constant, we are in position to present our final PDFs.
Given the robustness of the fitting ansatz reconstruction with respect to $\numax$ and also $\alpha_s$, we choose this approach as our preferred one.
The central values of our lattice-extracted PDFs use matched ITDs in the range $\nu\in[0,5.2]$ ($\zmax/a=8$), which fall on a universal curve, i.e.\ $Q(\nu,\mu^2)$ that can be obtained from different combinations of $(P_3,z)$ are compatible with each other within statistical uncertainties.

Apart from statistical errors of thus defined PDFs, we also add systematic uncertainties.
First, we consider the uncertainty related to the range of Ioffe times to be taken in the reconstruction procedure, $\Delta\zmax$. For each distribution $q$, we use a conservative definition, that is
\begin{equation}
\Delta\zmax(x) = \frac{|q_{\zmax/a=12}(x)-q_{\zmax/a=4}(x)|}{2}.
\end{equation}
Second, we consider the uncertainty from the choice of $\alpha_s$, $\Delta\alpha_s(x)$:
\begin{equation}
\Delta\alpha_s(x) = |q_{\alpha_s/\pi=0.129}(x)-q_{\alpha_s/\pi=0.1}(x)|.
\end{equation}
These two are added to the statistical error and can be considered as the quantified systematics of our result.

In addition, there are further systematic effects in the computation, that cannot be quantified at the present stage.
The latter will be subject to extensive follow-up work and will require additional numerical computations and further theoretical developments.
An extensive discussion of these effects is given in Ref.~\cite{Cichy:2018mum} and below, we comment on the most relevant points and we follow the strategy of Ref.~\cite{Cichy:2019ebf} of assuming plausible magnitudes of the considered effects, as percentages of ITDs values for different Ioffe times. We use estimates corresponding to scenario labeled S2 in Ref.~\cite{Cichy:2019ebf}, considered to be the most realistic one.

Discretization effects are an obvious source of systematics in lattice calculations.
Our results were obtained at a single value of the lattice spacing and thus, they are potentially contaminated by these effects.
To eliminate this uncertainty, simulations at preferably at least two additional lattice spacings are required.
Before this is done, in a longer time perspective due to the large cost of such simulations at the physical point, we assume that cutoff effects in our present data can be up to 20\%.
This number is rather conservative.
One argument to support this claim is that larger discretization effects would inevitably lead to the violation of the continuum dispersion relation, $E^2=P_3^2+m_N^2$, where $m_N$ is the nucleon mass.
Meanwhile, the dispersion relation was tested in Ref.~\cite{Alexandrou:2019lfo} and no deviations from the expected continuum behavior were found.
Additionally, related computations of moments of unpolarized PDFs by different groups (see e.g.\ Ref.~\cite{Constantinou:2014tga}) found deviations of $\mathcal{O}(5-15\%)$ between lattice results at lattice spacings similar to our and the continuum value.

\begin{figure*}[t!]
\begin{center}
    \includegraphics[width=0.9\textwidth]{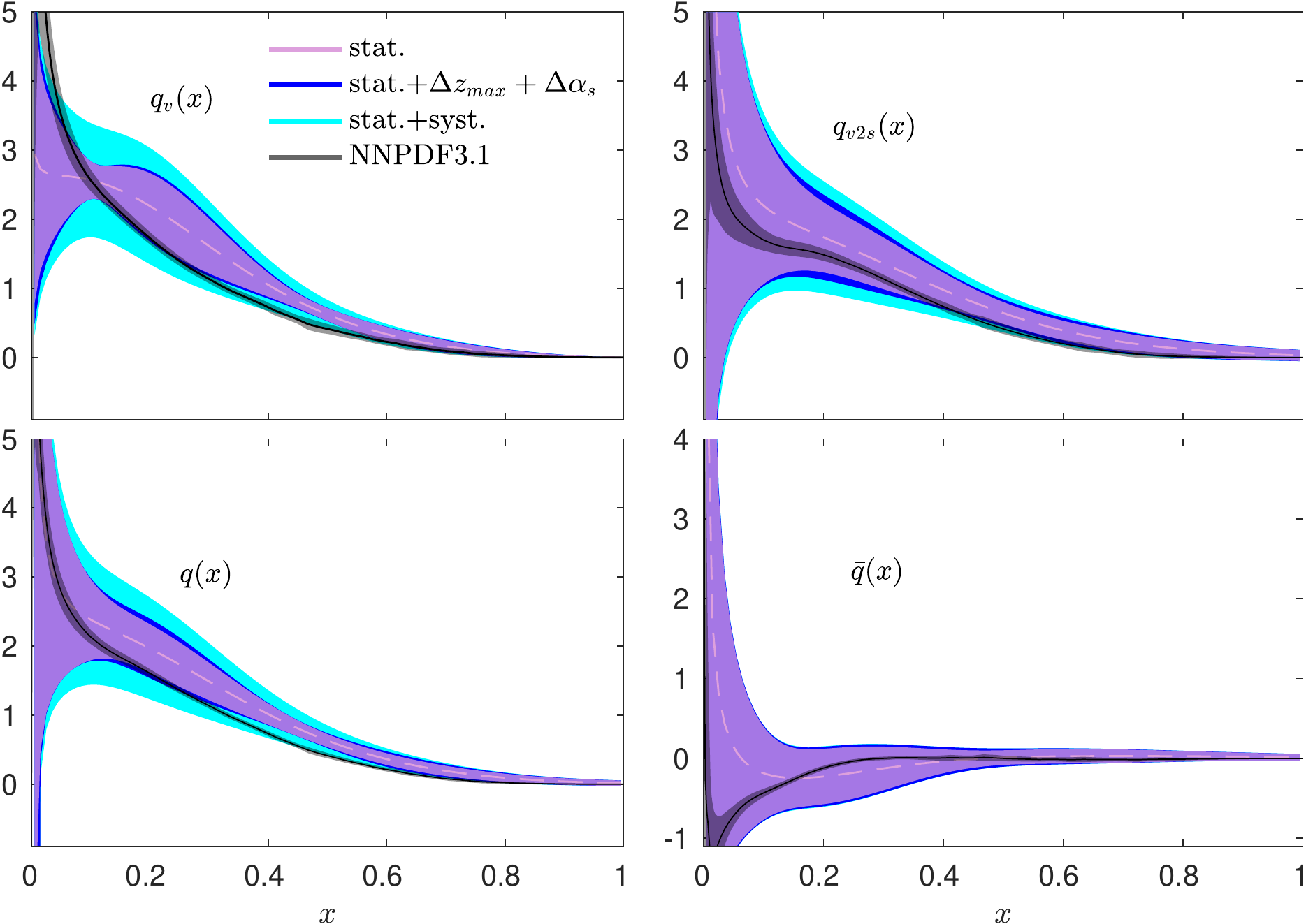}
\end{center}
\vspace*{-0.5cm} 
\caption{Final unpolarized PDFs extracted from the lattice (fitting ansatz reconstruction) and compared to global fits of NNPDF \cite{Ball:2017nwa} (solid black line and dark grey band). Shown distributions: valence ($q_v$; upper left), valence + 2 sea ($q_{v2s}=q_v+2\bar{q}$; upper right), full ($q=q_v+\bar{q}$; lower left) and sea ($\bar{q}$; lower right). The range of Ioffe times in the reconstruction is up to $\numax\approx5.2$ ($\zmax/a=8$), $\alpha_s/\pi\approx0.129$. The central value of each PDF is shown with dashed pink line and the bands represent the statistical uncertainty (purple), the latter with added uncertainty due to the choice of $\numax$ and $\alpha_s$ (blue) and the total error additionally with estimated uncertainties related to cutoff effects, FVE, excited states contamination, truncation and higher-twist effects (cyan) -- see text for more details. }
\label{fig:final}
\end{figure*}

The next source of systematics may be finite volume effects (FVE).
Before computations at additional lattice volumes are performed at the physical point to explicitly test them, we rely again on earlier studies, which typically find $\mathcal{O}(1-5\%)$ effects in related observables, provided that $m_\pi L\gtrsim 3$. 
In our case, $m_\pi L\approx 3$.
However, as pointed out in Ref.~\cite{Briceno:2018lfj} based on calculations in a toy scalar model, FVE in matrix element computations with spatially extended operators may be enhanced and the relevant parameter may be $m_N (L-z)$.
Since the nucleon mass is much larger than the pion mass, it would effectively not lead to any enhancement of FVE.
The worst plausible case is if in QCD the parameter controlling FVE becomes $m_\pi (L-z)$.
However, even then, the values of $z$ that are actually used should not lead to severe FVE.
This was confirmed for the $z$-dependent renormalization functions used in the non-perturbative renormalization of quasi-PDFs in Ref.~\cite{Alexandrou:2019lfo}.
Overall, to remain conservative, we allow for 5\% FVE in our hypothetical systematic error budget. 

Further systematic effects can result from contamination of the signal for the nucleon by excited states with the same quantum numbers.
For this uncertainty, investigated in great detail for the matrix elements used in this work, we rely on the conclusion of Ref.~\cite{Alexandrou:2019lfo}, where excited states suppression was found within statistical errors of the results.
Thus, we take this kind of systematics to be 10\% for all ITDs, i.e.\ slightly larger than the attained statistical precision.

The perturbative ingredient of our computation is subject to truncation effects.
These need to be investigated by calculating at least the two-loop matching.
Until this is carried out, the magnitude of this systematic effect is unknown. 
As we have demonstrated above, changing the $\alpha_s$ value for the one-loop formula does not lead to large changes of the PDFs.
However, the neglected higher-order effects may still be sizable.
The matching to light-cone ITDs, i.e.\ the factorization of the pseudo-ITD into its light-cone counterpart and a perturbative coefficient is also subject to HTE of $\mathcal{O}(z^2\LambdaQCD^2)$.
These effects, again, need to be further investigated with dedicated calculations, but they are not expected to be overwhelming with the values of $z$ that are included in the analysis and in view of the rather mild dependence on $\zmax$ that we find in this work.
Overall, for points discussed in this paragraph, we conservatively attribute 20\% as their potential size.

Our final PDFs are shown in Fig.~\ref{fig:final} and compared to global fits of NNPDF \cite{Ball:2017nwa}.
We show 3 kinds of error bands.
The purple one (the most narrow) is exclusively the statistical error of our results.
The systematic uncertainties, discussed above, enter in the blue band (quantified systematics from varying the range of Ioffe times and the value of the strong coupling) and in the cyan band (conservatively estimated errors from cutoff effects, FVE, excited states contamination, truncation and HTE).
The total uncertainty combines all the separate sources thereof in quadrature.

For all distributions, we find very good agreement with the corresponding phenomenological curve already within statistical errors, while the total error accounts for the remaining small discrepancies in certain regions of $x$ in $q_v$ and $q$ (around $x=0.5$ and also $x\lesssim0.05$ for $q_v$).
This gives confidence in the estimates of unquantified systematics, but we emphasize that much work is needed to properly quantify these effects.

The agreement of our final PDFs with NNPDF is striking and shows that a lattice extraction of the full $x$-dependence of PDFs is feasible.
It allows us also to draw conclusions about the reliability of such an extraction in different regions of $x$.
The most important numerical contributions to PDFs come from regions in Ioffe time where ITDs are large and where the Fourier transform cosine/sine weights are maximal.
In practice, this means relatively largest contributions from rather small Ioffe times.
For $q_v$, only this region has both large ITDs (real part) and a large weight.
For $q_{v2s}$, the imaginary part of ITDs increases linearly until $\nu\approx3$ and then stays approximately constant until the largest achieved Ioffe times and the latter fact combined with the periodicity of the sine function numerically suppresses the contributions from this region.
Overall, the large-$x$ region is observed to be almost insensitive to the range of Ioffe times when varying $\numax$.
Thus, it is clear that the large-$x$ part ($x\gtrsim0.6$) is reconstructed rather robustly for all distributions.
Moreover, one can argue that some other sources of systematics are expected to be small.
For instance, since small-$z$ contributions are observed to give largest numerical contributions, $\mathcal{O}(z^2\LambdaQCD^2)$ HTE are probably negligible, there are definitely no enhanced FVE of the type discussed in Ref.~\cite{Briceno:2018lfj} and cutoff effects are likely small, since the calculated small-$z$ matrix elements do not differ much from the local ($z=0$) ones, for which automatic $\mathcal{O}(a)$-improvement holds in our setup.
In the intermediate range of $x$, $x\approx0.2-0.6$, the uncertainties tend to increase and the one from varying the Ioffe time range becomes non-negligible (as wider range of Ioffe times becomes numerically important, i.e.\ the cosine/sine weights are varying more slowly, due to their argument being $x\nu$), even if it is still subleading.
The reconstructed PDFs are less constrained also due to part of the Ioffe time dependence of ITDs missing.
The latter becomes especially important in the low-$x$ region ($x\lesssim0.2$), where the total error becomes very large, particularly in distributions involving the imaginary part of ITDs that decays more slowly in Ioffe time than the real part.

\begin{figure}[h!]
\begin{center}
    \includegraphics[width=0.48\textwidth]{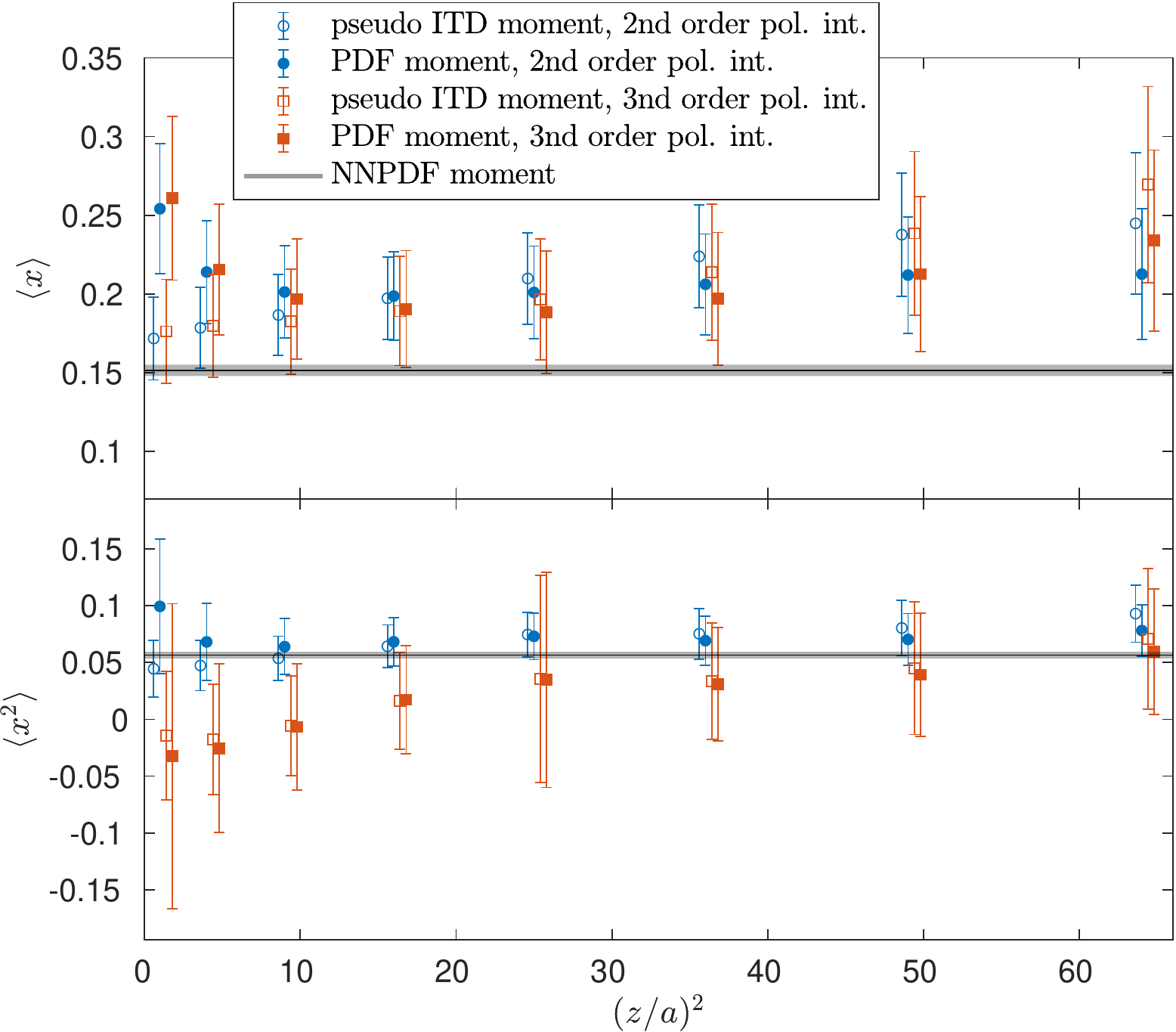}
\end{center}
\vspace*{-0.5cm} 
\caption{First (top) and second (bottom) moments of pseudo-PDFs (open symbols) and light-cone PDFs ($\MSb$ scheme at $\mu=2$ GeV; closed symbols), coming from second (blue) and third order (red) polynomial fits to the Ioffe time dependence of ITDs at fixed $z^2$.}
\label{fig:mom12}
\end{figure}

\begin{figure}[h!]
\begin{center}
    \includegraphics[width=0.48\textwidth]{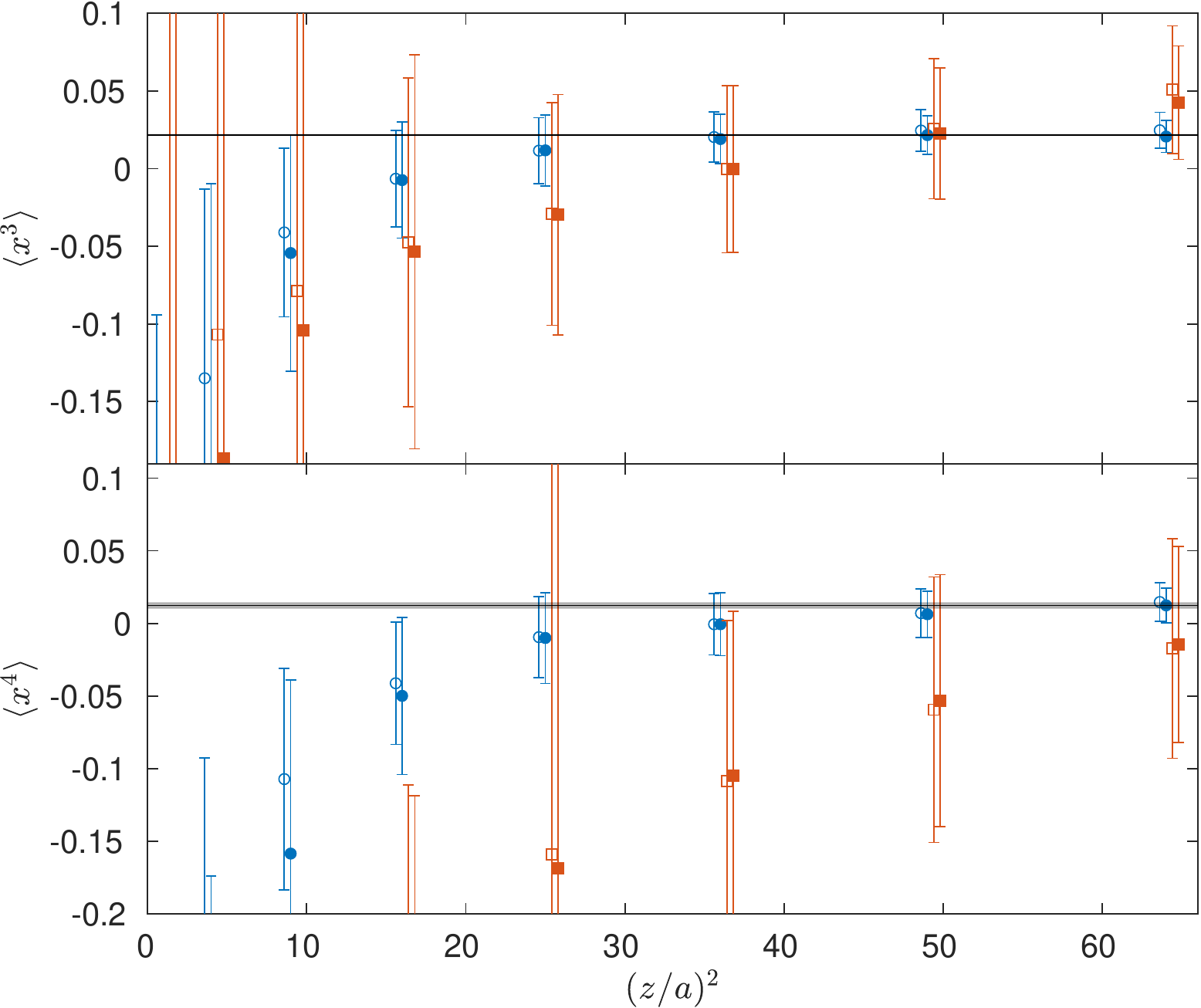}
\end{center}
\vspace*{-0.5cm} 
\caption{The same as Fig.~\ref{fig:mom12}, but for the third (top) and fourth (moments).}
\label{fig:mom34}
\end{figure}

\subsection{Moments analysis}
\label{sec:moments}
In this section, we use the polynomial fits to reduced ITDs at fixed $z^2$ to directly extract moments of pseudo-PDFs and relate them to moments of light-cone PDFs via the matching equation of Eq.~(\ref{eq:moments_matching}).
Then, we compare these to moments of our final reconstructed PDFs obtained from a numerical integration of the latter and to phenomenological ones, obtained from numerical integration of the NNPDF distribution \cite{Ball:2017nwa}.

In Fig.~\ref{fig:mom12}, we show our lowest two moments from the polynomial fits of second and third order.
The fits are performed independently at different discrete $z^2$ values, but should lead to the same moments up to HTE.
The observed independence of moments values within our precision confirms that HTE are indeed comparatively small at least up to $z^2\approx0.56$ fm$^2$, in line with our conclusions above at the level of final PDFs.
The results from the two considered polynomial orders are also consistent with each other, however, the third order fits are characterized by significantly larger errors.
In general, agreement between our extraction and NNPDF moments holds within 1-$\sigma$ for most cases, with a tendency to slightly too large values for $\langle x\rangle$.
Fig.~\ref{fig:mom34} illustrates that the third/fourth moments (and obviously all higher ones) cannot be meaningfully extracted at this level of precision.
The errors explode, particularly at small $z^2$ values, while the more precise second order polynomial fits at larger $z^2$ tend to reproduce the NNPDF values.

\begin{table}[h!]
\begin{center}
\renewcommand{\tabcolsep}{1pt}
\begin{tabular}{|c|c|c|c|c|c|c|}
\hline
 & NNPDF3.1 & \multicolumn{5}{|c|}{Lattice}\\
\hline 
& integration & \multicolumn{2}{|c|}{polynomial fits} & \multicolumn{3}{|c|}{integration}\\
\hline
 & Ref.~\cite{Ball:2017nwa} & 2nd ord. & 3rd ord. & nFT & BG & fits\\
\hline
$\langle x\rangle$ & 0.1515(36) & 0.199(29) & 0.190(37) & 0.23(5) & 0.26(6) & 0.231(35) \\
$\langle x^2\rangle$ & 0.0565(29) & 0.068(21) & 0.02(5) & 0.08(5) & 0.12(4) & 0.086(29) \\
$\langle x^3\rangle$ & 0.0217(6) & -0.01(4) & -0.05(13) & 0.04(4) & 0.07(4) &  0.043(22) \\
$\langle x^4\rangle$ & 0.01230(22) & -0.05(5) & -0.40(29) & 0.02(4) & 0.04(4) & 0.023(17) \\
\hline
\end{tabular}
\caption{Comparison of unpolarized PDF moments obtained from numerical integration of the NNPDF distribution \cite{Ball:2017nwa} with lattice evaluations, from second/third order polynomial fits to reduced ITDs at fixed $z^2\approx0.14$ fm$^2$ and from numerical integration of the final distributions from our three reconstruction approaches (with $\numax\approx5.2$ ($\zmax/a=8$), $\alpha_s/\pi\approx0.129$, $\MSb$ scheme at $\mu=2$ GeV).}
\label{tab:moments}
\end{center}
\end{table}

In Tab.~\ref{tab:moments}, we compare numbers obtained from the polynomial fits at a selected value of $z^2\approx0.14$ fm$^2$ and from numerical integration of our final PDFs, from our three reconstruction approaches.
Within our uncertainties, all the reported moments are consistent with one another.
All NNPDF moments are rather well-reproduced, although the uncertainties are relatively large.
The value of $\langle x \rangle$ from the integration of the fitting ansatz reconstructed PDFs is around 2-$\sigma$ above the phenomenological value, which results predominantly from the systematically too high values of the final PDFs at intermediate momentum fractions.
Other reconstruction methods lead to similar values of the moments, but their uncertainties are slightly larger than from the fitting ansatz reconstruction.
It is, however, reassuring to observe the agreement of the latter with NNPDF moments, given that they are comparatively more precise with respect to the ones from polynomial fits and statistically meaningful.

\section{Summary}
\label{sec:summary}
In this work, we used the pseudo-distribution approach to calculate the $x$-dependence of the unpolarized parton distribution functions of the nucleon.
The method relies on computations of spatial correlations between boosted nucleon states.
The resulting matrix elements are then used to form appropriate ratios that cancel the existing logarithmic and power-like divergences, defining pseudo-distributions in Ioffe time (Ioffe time distributions or ITDs).
Pseudo-ITDs are then matched to their light-cone counterparts, objects containing physical information and Fourier-conjugate to PDFs.
The step of translating the light-cone ITDs to PDFs is highly non-trivial due to the difficulty of obtaining the full Ioffe-time dependence of the former.
Thus, it is subject to an inverse problem and advanced reconstruction methods are needed instead of a simple Fourier transform.

We used a robust lattice setup of maximally twisted mass fermions to compute the bare matrix elements at the physical pion mass and with a relatively fine lattice spacing and large lattice volume that should lead to at most modest discretization and finite volume effects.
Having computed pseudo-ITDs, we performed the matching procedure to go to the light cone and we reconstructed the $x$-distributions using 3 methods.
As expected, the naive Fourier transform does not lead to robust results.
The Backus-Gilbert methods offers one solution to the inverse problem by assuming that the reconstructed distribution should have minimal variance with respect to the statistical variation of the data.
We found that with large range of the Ioffe-time dependence missing, this criterion is not good enough to reconstruct PDFs.
However, with this range being extended, the method gives results convergent with the ones from the third reconstruction approach.
The latter assumes a functional form of the light-cone PDFs, as done in phenomenological analyses to extract PDFs from global fits.
This assumption, similarly as the one in the Backus-Gilbert method, regulates the inverse problem and leads, in practice, to well-behaved and robust PDFs.
We checked this robustness by investigating the dependence of the reconstructed distributions with respect to the range of included Ioffe times and also the value of the strong coupling constant used in the matching.
We found that the large-$x$ region of PDFs is insensitive to this range, while at smaller $x$, ITDs at large Ioffe time lead to variations of the final PDFs only within statistical uncertainties.
Obviously, the missing data at these large Ioffe times increase the error of PDF estimates particularly at low-$x$.
To obtain more precise results there, ITDs need to be computed at larger nucleon boosts.
This presents a practical problem for LQCD, since the signal-to-noise ratio in the computation of matrix elements is quickly decaying with each additional unit of momentum.
Thus, it is natural that it will be difficult to reliably extract the very low-$x$ region.
Nevertheless, a large part of the $x$-dependence is possible to determine on the lattice with relatively modest computational resources.

Despite the optimistic results obtained here, it needs to be bore in mind that fully robust lattice-extracted PDFs need to have all relevant sources of systematic effects quantified.
Despite the overall excellent agreement with phenomenological distributions, there are intermediate stages where some tension with phenomenology is observed, such as the tendency that $\langle x\rangle$ is too large (at the level of 1-$\sigma$ to 2-$\sigma$; observed also in the study of Ref.~\cite{Joo:2020spy}) or the apparently too small/large values of the real/imaginary part of matched ITDs at intermediate and large Ioffe times (with similar tension).
A reliable quantification of all systematic effects will need a significant load of work in the next few years, involving additional computations on the lattice as well as theoretical developments.
Some sources of systematics can be quantified in a straightforward manner by repeating the procedure used in this work for additional ensembles of gauge field configurations, in particular at finer lattice spacings and larger volumes.
This will allow us to address, respectively, discretization and finite volume effects.
For other kinds of effects, proper strategies of addressing them need to be devised.
Truncation effects can only be properly quantified after derivation of higher-loop matching.
Higher-twist effects, in turn, can, in principle, be accessed numerically by computing proper matrix elements of more complicated operators, but analytical insight can be invaluable as well.

Thus, the field of extracting $x$-dependent PDFs, as well as other partonic functions, still needs a lot of progress.
However, it is clearly reassuring that the quality of our present results is already very satisfactory.
Only a few years ago, lattice data for PDFs were limited to only the lowest two or three moments, without realistic perspectives of reconstructing the $x$-dependence.
The progress induced by the seminal paper of Ji proposing how to extract the latter and subsequent alternative proposals, in particular the one we used in this work, change the prospects of this field to a huge extent.
As demonstrated in this work, the PDFs can indeed be extracted \emph{directly from first principles}, i.e.\ from the QCD Lagrangian and already now this can be done with both qualitative and quantitative agreement with global fits.
The latter requires additional work to be fully established and at this stage, we resorted to plausible hypotheses about the size of some systematic effects.

Given the success of this program for unpolarized PDFs, obvious directions for the future, apart from the discussed analysis of systematics, is to extend the work to polarized PDFs and other kinds of structure functions, in particular generalized parton distributions (GPDs) and transverse-momentum-dependent parton distributions (TMDs), as well as to singlet distributions.
All of these directions are challenging, but offer an unprecedented opportunity of having crucial insights for the partonic structure of the nucleon, relevant for its deeper understanding both at the theoretical and experimental level.

\begin{acknowledgements}
K.C.\ thanks Savvas Zafeiropoulos for many interesting discussions about the pseudo-distribution approach at different stages of evolution of this method.
M.B., K.C. and A.S.\ are supported by the National Science Centre (Poland) grant SONATA BIS no.\ 2016/22/E/ST2/00013. 
M.C.\ acknowledges financial support by the U.S. Department of Energy, Office of Nuclear Physics, Early Career Award under Grant No.\ DE-SC0020405. 
This research was supported in part by PLGrid Infrastructure (Prometheus supercomputer at AGH Cyfronet in Cracow).
Computations were also partially performed at the Poznan Supercomputing and Networking Center (Eagle supercomputer), the Interdisciplinary Centre for Mathematical and Computational Modelling of the Warsaw University (Okeanos supercomputer) and at the Academic Computer Centre in Gda\'nsk (Tryton supercomputer). \end{acknowledgements}


\bibliography{references.bib}

\end{document}